\documentclass[twocolumn]{aastex}

\usepackage{amsmath}
\usepackage{bm}% bold math
\usepackage{epsfig}
\usepackage{graphicx}% Include figure files
\usepackage{dcolumn}% Align table columns on decimal point
\usepackage{xcolor}
\usepackage{subfigure}
\usepackage{amssymb}
\usepackage{amsfonts}
\usepackage{color}
\usepackage{epsf}
\usepackage{float}
\usepackage{latexsym}
\usepackage{mathrsfs}
\usepackage{tikz}
\usepackage{verbatim}
\usepackage[utf8]{inputenc}
\usepackage{url}
\newcommand{\Mpc}{\ensuremath{\,{\rm Mpc}}}
\newcommand{\Gpc}{\ensuremath{\,{\rm Gpc}}}

\newcommand{\MHz}{\ensuremath{\, {\rm MHz}}}

\renewcommand{\deg}{\ensuremath{\,{\rm deg}}}

\def\ba{\begin{align}}
\def\ea{\end{align}}
\def\be{\begin{equation}}
\def\ee{\end{equation}}
\def\bea{\begin{eqnarray}}
\def\eea{\end{eqnarray}}

\begin{document}
\title{Stages of Reionization as Revealed by the Minkowski Functionals}
% -- percolation and neutral islands in the late stage 

\author{Zhaoting Chen}
\affiliation{Key Laboratory for Computational Astrophysics, National Astronomical Observatories, Chinese Academy of Sciences, Beijing 100101, China}
\affiliation{Department of Astronomy, University of Science and Technology of China, Hefei , China}

\author{Yidong Xu}
\affiliation{Key Laboratory for Computational Astrophysics, National Astronomical Observatories, Chinese Academy of Sciences, Beijing 100101, China}

\author{Yougang Wang}
\affiliation{Key Laboratory for Computational Astrophysics, National Astronomical Observatories, Chinese Academy of Sciences, Beijing 100101, China}

\author{Xuelei Chen}
\correspondingauthor{Yidong Xu, Xuelei Chen}
\affiliation{Key Laboratory for Computational Astrophysics, National Astronomical Observatories, Chinese Academy of Sciences, Beijing 100101, China}
\affiliation{School of Astronomy and Space Science, University of Chinese Academy of Sciences, Beijing 100049, China}
\affiliation{Center of High Energy Physics, Peking University, Beijing 100871, China}
\email{xuyd@nao.cas.cn, xuelei@cosmology.bao.ac.cn}

\begin{abstract}
We study the evolution of the hydrogen ionization field during the epoch of reionization (EoR) using 
 semi-numerical simulations. By calculating the Minkowski functionals (MFs) of the 21cm brightness temperature 
 field, which provides a quantitative description of topology of the neutral and ionized regions, 
 we find that the reionization process can be divided into five stages, each with different topological structures, 
corresponding to isolated ionized regions (``bubbles"); connected ionized region (``ionized fibers") which 
percolate through the cosmic volume;  a sponge-like configuration with intertwined neutral and ionized regions after the 
overlap of bubbles; connected neutral regions (``neutral fibers") when the ionized region begin to surround the remaining 
neutral region; and isolated neutral regions (``islands"), before the whole space is ionized except for the rare 
dense clumps in galaxies and minihalos and the reionization process is completed. We use the MFs and the size statistics
on the ionized or neutral regions to distinguish these different stages, and find the neutral fractions at which the 
transition of the different stages occur.  At the later stages of reionization the neutral regions are more isolated than the 
ionized regions, this is the motivation for the island model description of reionization. We compare the late stage topological
evolution in the island model and the bubble model, showing that in the island model the neutral fibers are more easily 
broken into multiple pieces due to the ionizing background.  
\end{abstract}


\section{introduction}
The epoch of reionization (EoR) is a complex and information-rich stage in the history of our
Universe, during which the first galaxies formed and the intergalactic medium (IGM) transits from neutral to ionized
(e.g. \citealt{2018PhR...780....1D}).
The Lyman-$\alpha$ absorption spectra of high redshift quasars show that reionization is nearly complete 
at $z\approx6$ \citep{Fan:2006dp}, while the latest cosmic microwave background (CMB) data from the Planck satellite 
sets the average redshift of EoR to be $z_{re}=7.82$ \citep{Akrami:2018vks}. 
Recently, the Experiment to Detect the Global Epoch of Reionization Signature (EDGES) claims to 
detect a strong global 21cm absorption signal at redshift $z\sim 18$ corresponding to the formation of the first luminous 
objects \citep{2018Natur.555...67B}, though the claimed signal is too strong to be consistent with standard cosmological 
model, and this result still needs further confirmation from more experiments. 
Besides the basic timing of reionization, however, more details
about this process, such as the nature and properties of the ionizing sources and the scale and morphology 
of the ionization field are still very uncertain. Using the 21 cm transition of neutral hydrogen as the tracer of 
the neutral IGM, various experiments
are built or being designed in order to detect the 21 cm signals from the EoR, including the current GMRT \citep{Paciga:2010yy}, 
MWA \citep{Tingay:2013ypa}, PAPER \citep{Parsons:2009in}, and LOFAR \citep{vanHaarlem:2013dsa}, 
and the upcoming HERA \citep{DeBoer:2016tnn} and SKA-low \citep{Maartens:2015mra}.
The 21 cm power spectrum may be the most comprehensively studied statistics, and
it is also probably the first statistical signal to be measured in the near future.
Nevertheless, much more information is expected to be extracted from other 21 cm statistics beyond 
the power spectrum, as the reionization process involves various non-linear processes,
and the ionization field is expected to be highly non-Gaussian.
Indeed, evolutionary features in the skewness (e.g. \citealt{Wyithe:2007if}), kurtosis (e.g. \citealt{Kittiwisit:2017jia}), and bispectrum (e.g. \citealt{Shimabukuro:2015iqa}) of the 21 cm brightness fields, among others, have been explored.

According to our current understanding, reionization began with the formation of first stars and galaxies
in high density regions, which photoionized the gas in the surrounding regions and form individual ionized bubbles. 
As more and more galaxies form, the bubbles grow in size and number, at some point they
start to connect with each other, going through a percolation process, and eventually the ionized regions overlap with 
each other and most of them are connected as a whole. The remaining neutral regions are then one-by-one 
cut off from each other and become isolated islands. The reionization process 
is completed with the shrinking and disappearing of these islands.

The  widely adopted ``bubble model'' assumed isolated-bubbles topology for the early stage of 
reionization \citep{Furlanetto:2004nh}, and several semi-numerical simulation programs have been 
developed based on this idea \citep{Zahn:2006sg,Mesinger:2010ne}). 
It is important to know the realm in which the isolated-bubbles topology applies, before such simulations
are used for interpreting the upcoming observational data. 
Recently, it has been recognized that the bubble overlap happens at fairly high neutral 
fractions \citep{Xu:2013npa,Furlanetto:2015hqp}, casting doubts on the applicable range of the models based on the
idea of isolated bubbles, as its basic premise is invalid except for a very brief periods. 
At a later stage, when most of the space have been reionized, the neutral regions become isolated, 
and one could then treat them as neutral islands \citep{Xu:2013npa}, but 
again we need to know the point at which the topological phase transition happened and the 
model based on the island picture can apply. 

The two-dimensional and three-dimensional topology of the ionized regions during the EoR have been 
investigated in a number of 
researches \citep{Gleser:2006su,2008ApJ...675....8L,2014JKAS...47...49H,Wang:2015dna} 
\citep{2018JCAP...10..011K,Bag:2018zon,Bag2019MNRAS}.
While the neutral fraction, matter density and spin temperature of the hydrogen atoms may all affect 
the 21cm signal,  \citet{Yoshiura:2016nux} demonstrated that the MFs of the 
21cm field is determined primarily by the neutral fraction, so that it provides a very powerful tool to 
study the topological evolution process during reionization. 

Here we aim to quantify the topological structure  of the 21 cm brightness field in terms of the MFs, 
and provide a full description of the evolutionary features of the 21 cm signal through the reionization process.
Using the Minkowski functionals and the statistics of ionized and neutral regions, we try to gain an understanding of
how the topology of the neutral and ionized regions evolve.

In Section 2,  we briefly introduce the excursion set theory of reionization, and 
the semi-numerical simulations, on which our topological analysis is based. The computation of MFs 
is given in Section 3.  The evolutionary features in the topology
of the 21 cm brightness field are presented in Section 4, with each subsection dedicated to a stage of EoR. 
Section 5 compares the topology evolution in the 21cmFAST and the IslandFAST. 
We then discuss the feasibility to measure the MFs with SKA in Section 6, 
before summarizing the main results in Section 7.

\section{Semi-numerical simulations}
In order to generate the ionization field and the corresponding 21 cm brightness temperature map during reionization,
we use the semi-numerical simulations of the 21cmFAST \citep{Mesinger:2010ne} program as well as 
the IslandFAST\citep{Xu:2016nmi}, which is designed to have a more accurate description at the late stage of the reionization.  
The semi-numerical simulations are based on the excursion set theory of structure 
formation (\citealt{Bond:1990iw}, \citealt{Lacey:1993iv}, \citealt{Zentner:2006vw}), in which the formation of 
luminous objects are determined by the over-density on the corresponding mass scales, and the ionization of the region
is determined by the balance of ionizing photon supply and the number of baryons to be ionized--in the presence of recombination, 
each baryon may also require multiple photons to keep it ionized during the time being considered.  
The 21cmFAST program was developed based on the ``bubble model'' of reionization \citep{Furlanetto:2004nh}
which assumes an isolated-bubble topology of the ionization regions.
To obtain a more accurate description of the reionization process after percolation of the ionized regions, the ``island model''
\citep{Xu:2013npa} and the corresponding semi-numerical code IslandFAST \citep{Xu:2016nmi} were developed,
assuming a topology of isolated neutral islands in the ionization field.
Strictly speaking, the ``bubble model'' should be applied only before the connection of individual bubbles, 
while the ``island model'' should be used only after the isolation of the neutral patches, and during the mid-stage
of reionization when the percolation occurs, neither of these model is completely accurate. 
Within the framework of the excursion set theory, the watershed happens when the density barrier
crosses the coordinate origin on the $\delta-S$ plane.

Although the ionized bubbles are isolated only at the early stage of reionization, it was found that the statistical predictions from 
the 21cmFAST still have a reasonable agreement with full radiative transfer hydrodynamical simulations 
for most epochs during reionization \citep{Zahn:2006sg}. In this paper 
we will use the 21cmFAST to simulate the whole process of reionization, and analyze 
the topology of reionization. Using the MFs, we will identify the point
at which neutral regions are divided into discrete pieces, after that the neutral island gives a better 
description of the hydrogen distribution. For the late stage, we also perform the IslandFAST simulation, which 
begin with an initial state generated with 21cmFAST, but then evolve the reionization sequence using the 
IslandFAST rules.

We assume an ionizing efficiency parameter of $\zeta = 20$, and a minimum virial temperature of 
$10^4$ K for halos hosting ionizing sources as our fiducial model.
For the IslandFAST, an empirical model for the ionizing background is adopted, based on the
observed number density of Lyman limit systems near redshift 6 \citep{2010ApJ...721.1448S}.
The simulation box has comoving length of 1 Gpc on a side and $500^3$ cubic cells.

The EoR experiments measure the 21 cm brightness temperature, 
given by (e.g. \citealt{Furlanetto:2006jb}):
\begin{eqnarray}
T_b (z)&\approx& 27x_{\rm HI}(1+\delta_{m})\left(\frac{H}{{\rm d}v_{r}/{\rm d}r+H}\right)
\left(1-\frac{T_{\gamma}}{T_{\rm S}}\right) \nonumber \\
 &&\left(\frac{1+z}{10}\frac{0.15}{\Omega_{m}h^2}\right)^{1/2}\left(\frac{\Omega_{b}h^2}{0.023}\right)[{\rm mK}],
\end{eqnarray}
where $H$ is the Hubble parameter, $\Omega_{m}$ and $\Omega_{b}$ are the matter and 
baryon density parameters respectively, $T_{\gamma}$ is the CMB temperature and 
${\rm d}v_{r}/{\rm d}r$ is the radial gradient of the peculiar velocity. 
The topology of $T_b$ contains the three-dimensional information about the neutral fraction $x_{\rm HI}$, 
matter overdensity $\delta_{m}$, spin temperature $T_{\rm S}$, as well as the radio velocity gradient. 
The spin temperature is an weighted average of the gas kinetic temperature and the CMB temperature. 
At the cosmic dawn, when the first stars and galaxies formed, the spin temperature might be driven below the CMB temperature 
by the Wouthuysen-Field mechanism of the Lyman-$\alpha$ photons, producing an absorption 
trough \citep{Chen:2003gc}. However, for the major part of the epoch of reionization, especially when the ionization 
fraction became significant, the gas would have already been heated well above the CMB temperature, thus the tomographic
signal would be found 21cm emission. Below for simplicity  we assume 
$T_{\rm S} \gg T_{\gamma}$ in the analysis.
While the density fluctuations also contribute to $T_b$ fluctuations, the ionization can set 
the brightness directly to 0. Therefore, the features in the MFs of $T_b$ 
during reionization are dominated by the $x_{\rm HI}$ distribution \citep{Yoshiura:2016nux}. 
We will focus on the topology of $T_b=0$ contours.

\section{Minkowski Functionals} 

To quantify the morphological structure of the reionization, we calculate the 
MFs \citep{Minkowski1903} of the $T_b$ field, for which there are only four independent ones in three 
dimensional space \citep{1997ApJ...482L...1S}.  A typical set of MFs in the three dimensional space is:
\be
V_0(\nu)=\frac{1}{V}\int_{V}{\rm d}^{3}x~ \Theta[u(\mathbf{x})-\nu\sigma]
\ee
\be
V_{1}(\nu)=\frac{1}{6V}\int_{\partial F_{\nu}}ds
\ee
\be
V_{2}(\nu)=\frac{1}{6\pi V}\int_{\partial F_{\nu}} {\rm d}s~ [\kappa_{1}(\mathbf{x})+\kappa_{2}(\mathbf{x})]
\ee
\be
V_{3}(\nu)=\frac{1}{4\pi V}\int_{\partial F_{\nu}} {\rm d}s ~\kappa_{1}(\mathbf{x})\kappa_{2}(\mathbf{x})
\ee
where $\Theta$ denotes the Heaviside step function, $u(\mathbf{x})$ is the random field being studied, $\sigma$ is the 
root mean square (rms) of the field, and $\nu=u_{\rm thr}/\sigma$ is the relative threshold. $V$ denotes the three dimensional 
space being studied, $x$ denotes the comoving coordinates, $\partial F_\nu$ denotes the contour surface formed by 
 $u(\vec{x})=\bar{u}+\nu\sigma$,  $\kappa_{1,2}$ are the principle curvatures of $\partial F_\nu$. 
 These MFs, in order, describes the volume, surface area, mean curvature and the Gaussian curvature of the excursion sets, 
   i.e. regions in which the field $u$ exceeds certain threshold. By the Gaussian integration theorem, $V_3$ is also  
the Euler characteristic, which is related to the genus by $g=1-\chi/2$, which describes the topology of the excursion set, with 
 $g V$ = (number of tunnels) - (number of isolated surfaces)+1.
 
We smooth the $T_b$ field with the simple average and a smoothing scale of $4^3$ cells (8 Mpc a side). 
The 21cm fluctuation $T_b/\sigma_{T_b}$ is used as the probe field,  and as we discussed above, 
 the $T_b=0$ contours  trace the ionization fronts. 
 During the reionization, at least in popular reionization models with soft ionizing photon sources, the 
 boundary between the ionized regions and neutral regions are sharply defined. At such boundary $T_b=0$, 
 thus we may watch for the $T_b=0$ point in the MF for the 
evolution of ionized regions. Other points in the MFs reflect also the change in the density of the neutral hydrogen field, though 
being affected by many factors it is more difficult to interpret.
 
 We use the Koenderink method \citep{Koenderink1984} to 
 calculate the MFs. This method obtain the principal curvature by calculating
 the first and second order partial derivatives of the field, and transforms the surface integrals to volume integrals for 
 computation using the Gaussian theorem.  
 
\begin{figure}[htbp]
\centering
\includegraphics[width=0.23\textwidth]{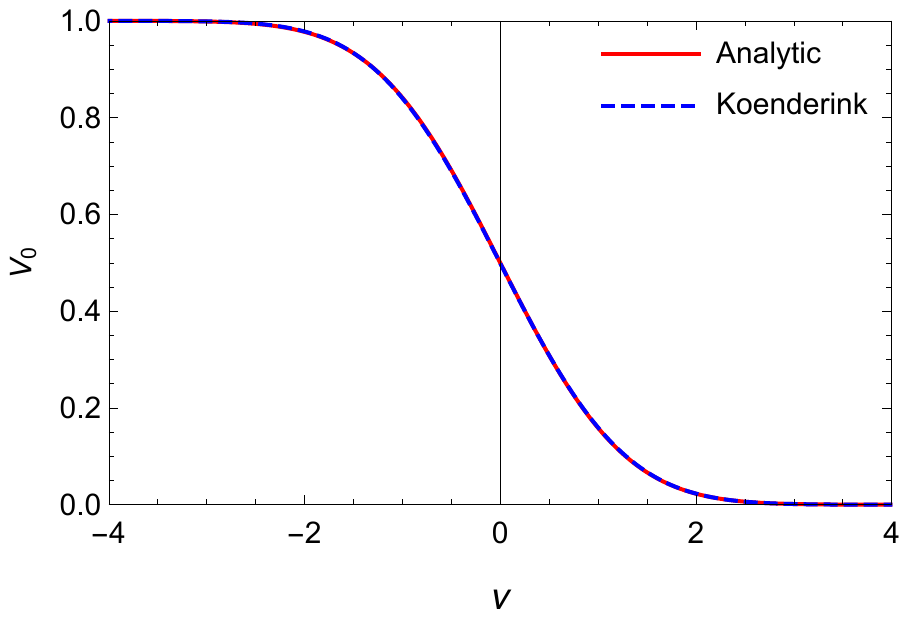}
\includegraphics[width=0.23\textwidth]{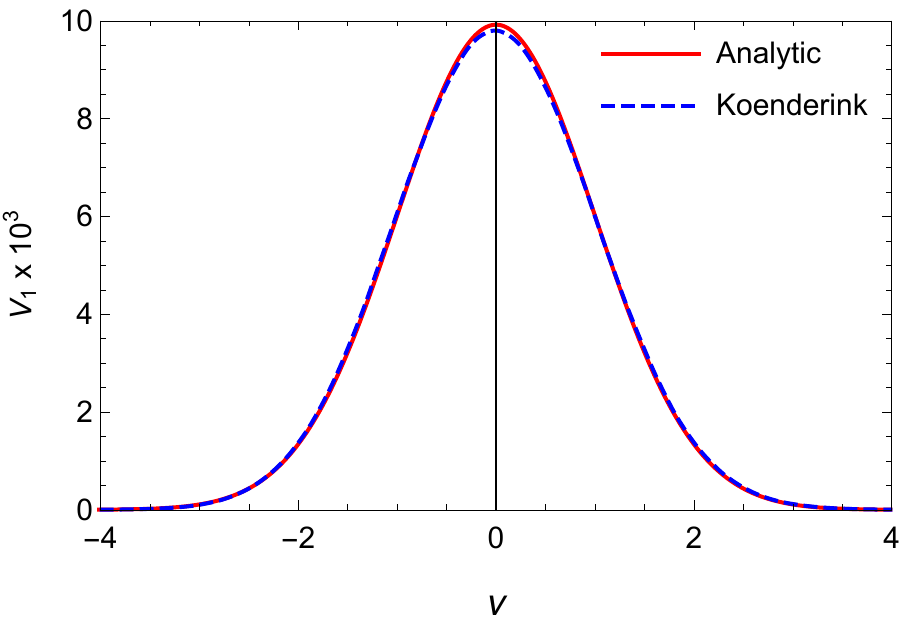}\\
\includegraphics[width=0.23\textwidth]{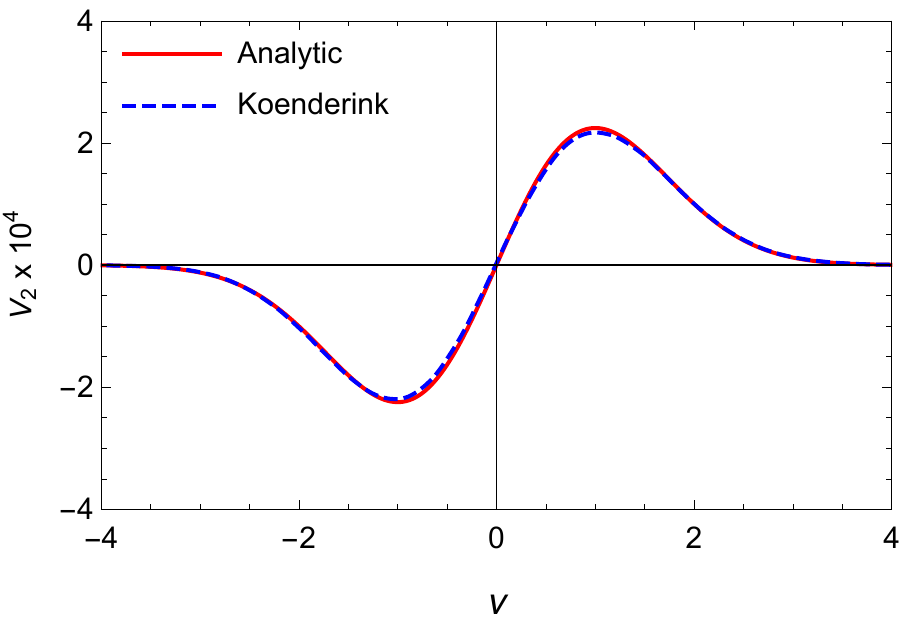}
\includegraphics[width=0.23\textwidth]{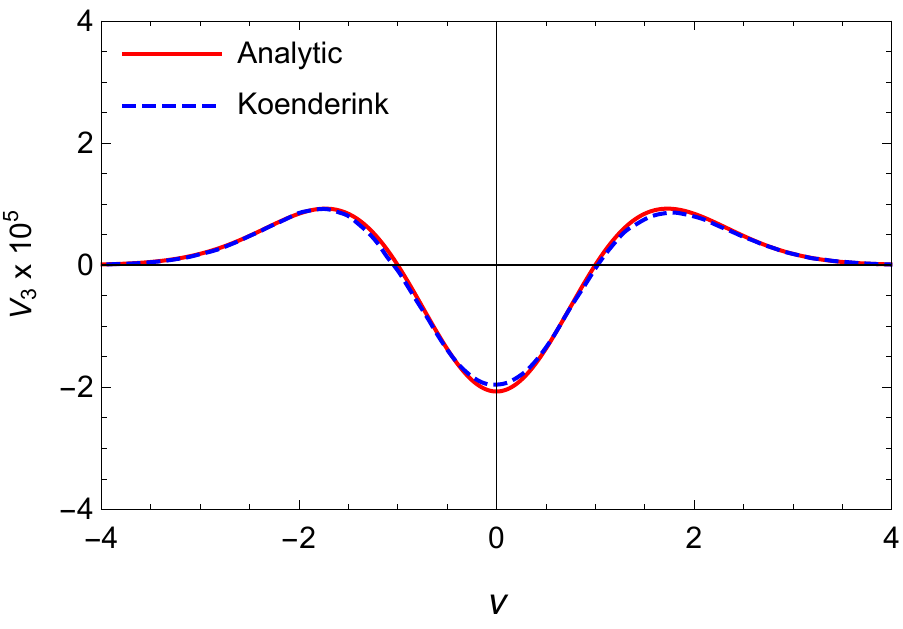}\\
\caption{MFs for a Gaussian random field with $\bar{u}=0$ and $\sigma=0.5$. Red solid line denotes analytical solution and blue solid line denotes the result of numerical algorithm, which are almost identical. The resolution in this case is $128^3$.}
\label{figGauss}
\end{figure}

Before moving on to measuring the simulation results, we first consider the case of random Gaussian field, 
which gives a good description of primordial density field of the large scale structure in the standard cosmology model.  
The analytical form of the MFs are known in this case:
 \begin{eqnarray}
 V_0&=&\frac{1}{2}-\frac{1}{2} \Phi(\frac{\nu}{\sqrt{2}})\\ 
 V_1&=&\frac{2}{3}\frac{\lambda}{\sqrt{2\pi}}e^{-\nu^2/2}\\
 V_2&=&\frac{2}{3}\frac{\lambda^2}{\sqrt{2\pi}} \nu e^{-\nu^2/2}\\
 V_3&=&\frac{\lambda^3}{\sqrt{2\pi}}(\nu^2-1) e^{-\nu^2/2}
 \end{eqnarray}
 where $\nu=\delta/\sigma$, $\lambda=\sqrt{|\xi^{''}(0)|/(2\pi\xi(0))}$ where $\xi(r)$ denotes the  correlation function at zero, 
 and $\Phi(x)=\frac{2}{\sqrt{\pi}} \int_0^x dt ~e^{-t^2}$ is the Gaussian error integral. We plot the MFs of a Gaussian 
random field in Fig.~\ref{figGauss} for reference, and use them to check our code for numerically measuring the MFs.

We can also gain some insights on the interpretation of the MFs from this plot.  The left end of the plot corresponds to 
extreme under-density,  so the contours  there would have very small total surface area ($V_1 \sim 0$), and as the surface
is oriented from high density to low density, it enclosed most of the volume ($V_0 \sim 1$). 
As the density increases, the enclosed volume fraction $V_0$ decreases, which drops to zero at the right end
for the  extremely large over-density fluctuations. During the same process, the area of the contour surface $V_1$ initially 
increases as we move to more probable field values, reaching a maximum at the mean value of $\nu=0$, then starts
decreasing again to smaller values at large over-densities.  The mean curvature $V_2$ at $\nu<0$ is negative due to the 
orientation of the contour surface and positive at $\nu>0$, crossing zero at the mean field value of $\nu=0$. 
It approaches zero at large $|\nu|$, as the number and contour surface of such extreme points rapidly decrease, though the 
curvature itself is large. The maximum and minimum of $V_2$ is reached near $\nu=1$ and $-1$ respectively. 
 However, $V_2$ is not frequently used in cosmological applications.
The $V_3>0$ peaks at a moderately large $|\nu|$ indicated that $g<0$, i.e. the number of tunnels smaller than the 
number of isolated surfaces, while the  case $V_3<0$ near $\nu=0$ is just the opposite, where we would have a ``sponge" 
topology with complex multi-connected iso-density surfaces.

\begin{figure*}[tbp!]
\centering
\includegraphics[width=\textwidth]{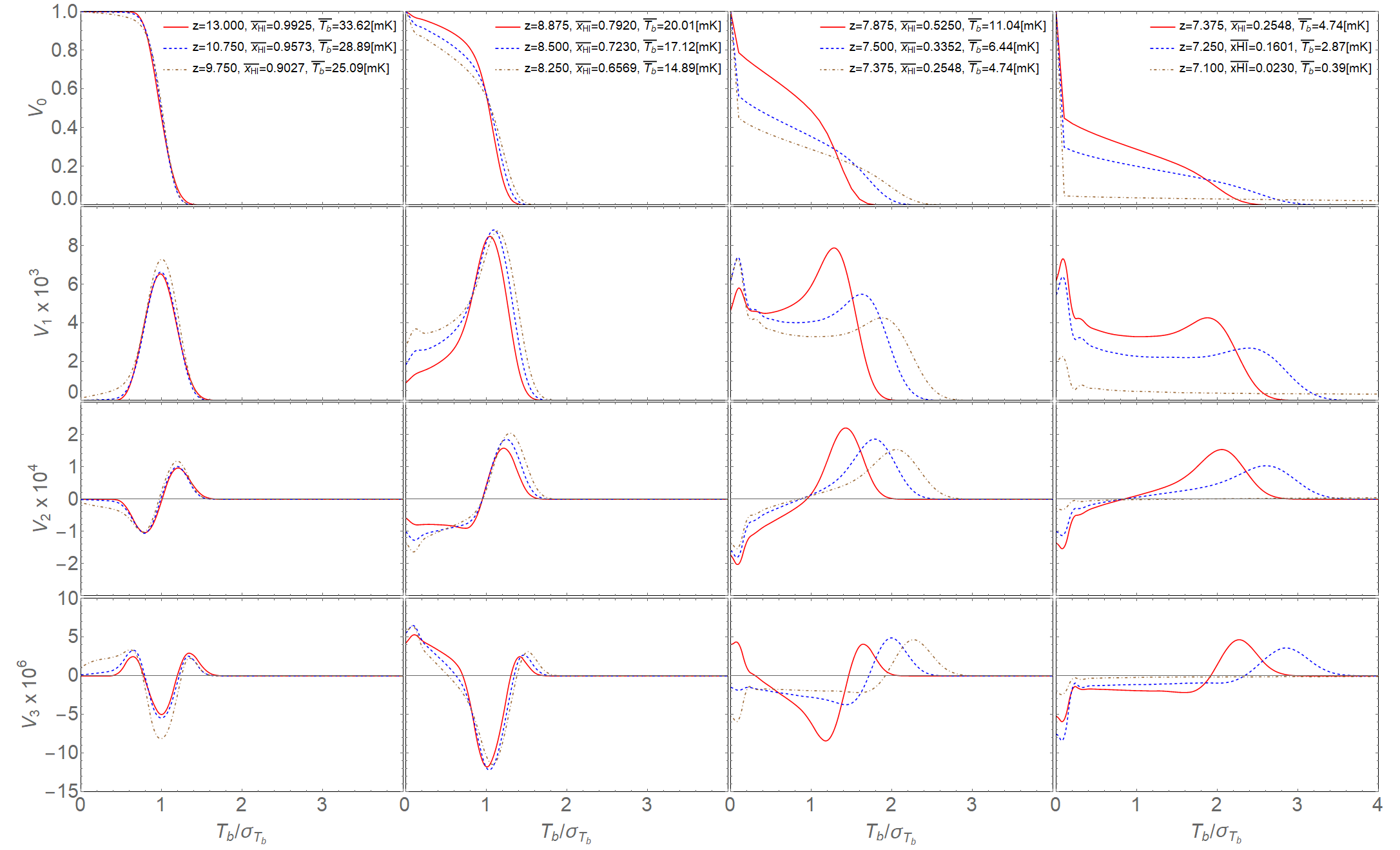}
\caption{The Minkowski functionals at different redshifts. From top to bottom are $V_0, V_1, V_2, V_3$ respectively, 
and from left to right, for each panel the MF at three redshifts (marked on the plot) are shown.}
\label{figMF}
\end{figure*}

\section{Stages of Reionization} 

In the bubble model, it is noted that the ionized bubbles first formed in high 
density peaks where the star formation rate is high, and as redshift decreases, 
more bubbles appear and existing bubbles grow in size, eventually  they would overlap with each other 
\citep{Furlanetto:2004nh}, then most of the space is reionized, with only a small fraction of low density regions 
remain as neutral islands, and afterwards they also disappeared quickly. 
After that, the hydrogen reionization is complete and the neutral hydrogen is only found within the high 
density clumps of galaxies or minihalos.  
However, it has been recognized recently that the percolation of the ionized bubbles could occur 
when the average ionized fraction is still very low \citep{Xu:2013npa,Furlanetto:2015hqp}, so strictly speaking, 
the {\it bona fide} isolated bubble picture is only valid for a very brief period.  

Armed with the  MFs, we can now study the history of reionization quantitatively. However, we shall use a modified form. 
We shall plot the $V_i$ not as a function of the temperature fluctuation normalized with the variance, 
$\delta_T = (T_b-\bar{T_b})/\sigma_{T_b}$, but simply as a dimensionless brightness temperature
$T_b/\sigma_{T_b}$. 
The advantage of using this is that the boundary between neutral and ionized regions have $T_b=0$, which we can easily 
mark, while in the conventional definition this point changes with $\bar{T_b}$.

The redshift evolution of the MFs are plotted in Fig.~\ref{figMF}, where the MFs $V_0,V_1,V_2,V_3$ are shown from top 
to bottom, the results at different redshifts are plotted on the four columns from left to right in descending order, and within 
each panel results for three redshifts are plotted.  

Using the MFs as a classification tool, we find that the reionization history 
can be divided into the following five stages according to morphology: 
\begin{itemize}
\item {\bf The Ionized Bubbles stage}.
Initially the MFs of the 21cm brightness temperature field are almost identical with the case of a 
Gaussian random field. A  small number of isolated ionized bubbles begin to appear in the volume at high peaks of the 
density field. 
\item {\bf The Ionized Fibers stage}. From $x_{\rm HI}\sim0.9$ to $x_{\rm HI}\sim0.7$,  more and more 
bubbles get connected to form a large fiber-like structure extending the whole simulation box. 
While such ionized structures do not look exactly as the originally envisioned bubbles, they can still be regarded as 
connected bubbles embedded in the neutral box.
\item {\bf The Sponge stage}. 
When $x_{\rm HI}$ drops below 0.7, we begin to see that most of the ionized regions get connected, 
with the neutral and ionized regions intertwined with each to form a sponge-like complex structure. 
\item{\bf The Neutral Fibers stage}. 
When the neutral fraction drops below 0.3, the neutral regions begin to look like a large fiber embedded in the 
ionized regions. This stage mirrors the ionized fiber stage before the sponge stage. 
\item {\bf The Neutral Islands stage}.
As reionization progresses further, the connections between the neutral regions get cut off. When $x_{\rm HI} < 0.16$,
the remaining large neutral regions become islands, and continue to shrink and disappear in the end. 
\end{itemize}

Below, we study the evolving MFs and the stages of reionization in detail. 
The transition neutral fraction between each stage and the corresponding features in MFs are summarized
in Table~\ref{table:1}.

\subsection{The Ionized Bubbles stage}
In Fig.~\ref{figMF}, the first column from left shows the earliest stage of the reionization process. The three curves
in the figure are for redshift $z=13$ (red solid line), $z= 10.75$ (blue dashed line), and $z=9.75$ (brown dash-dotted line),
and the corresponding neutral fractions are $x_{\rm HI}= 0.9975, 0.9573, 0.9025$ respectively (At the mass 
resolution considered here, the difference between the volume and mass-weighted neutral fractions are not large, we
shall use the volume neutral fraction in all discussions in this paper.)
The general impression on the MFs of this stage is that they are very close to the Gaussian case, with only minor deviations. 

At the beginning of reionization ($z=13$), almost all of the gas are neutral, and the 21 cm brightness field are 
dominated by the density fluctuations. The MFs of the $T_b$ field follow those of a Gaussian field.

As the ionized bubbles formed,  the neutral hydrogen field deviates from the Gaussian case. Inside the ionized regions
 $T_b=0$, and the boundary between the ionized region and neutral regions are sharp, at least for the soft 
 ionizing sources such as stars, so we  expect that the volume $V_0(T_b)$ should drop steeply 
 just above the point $T_b=0$, as a fraction of the volume 
 $x_i=1-x_{\rm HI}$ is now ionized and has $T_b=0$. It should rejoin the original Gaussian curve at a height of 
 $V_0=x_{\rm HI}$, if the brightness temperature field remain unchanged outside ionized region. Due to the effect of 
 smoothing, the actual drop of $V_0$ is not  so steep. In fact for $x_{\rm HI} \approx 0.96 $ case, the  difference from 
 the Gaussian curve is very small, perhaps because the bubbles are still smaller than the smoothing scale.
  But we do see that for the $x_{\rm HI} \approx 0.90$ case, the $V_0$ decreases at $T_b>0$ and  
  approaches the Gaussian curve at a value close to $x_{\rm HI}$. The same is true for 
  the surface area $V_1$, which for $x_{\rm HI}\approx 0.96$ 
  the change is unnoticeable, but for $x_{\rm HI} \approx 0.90$  the increase in $V_1$ is quite obvious. 
 
The nature of the ionized region can be revealed by $V_3$ which is related to the topology of the field.  
It has an increase at $T_b>0$, showing an excess of contour surfaces, corresponding to the 
formation of bubbles. The trough of $V_3$ 
at the mean brightness temperature also deepens, perhaps due to structure growth, but this is more complicated and 
not relevant to the main purpose of our study, so we will not discuss it further.

\begin{figure} [htbp]
\centering
\includegraphics[width=0.28\textwidth]{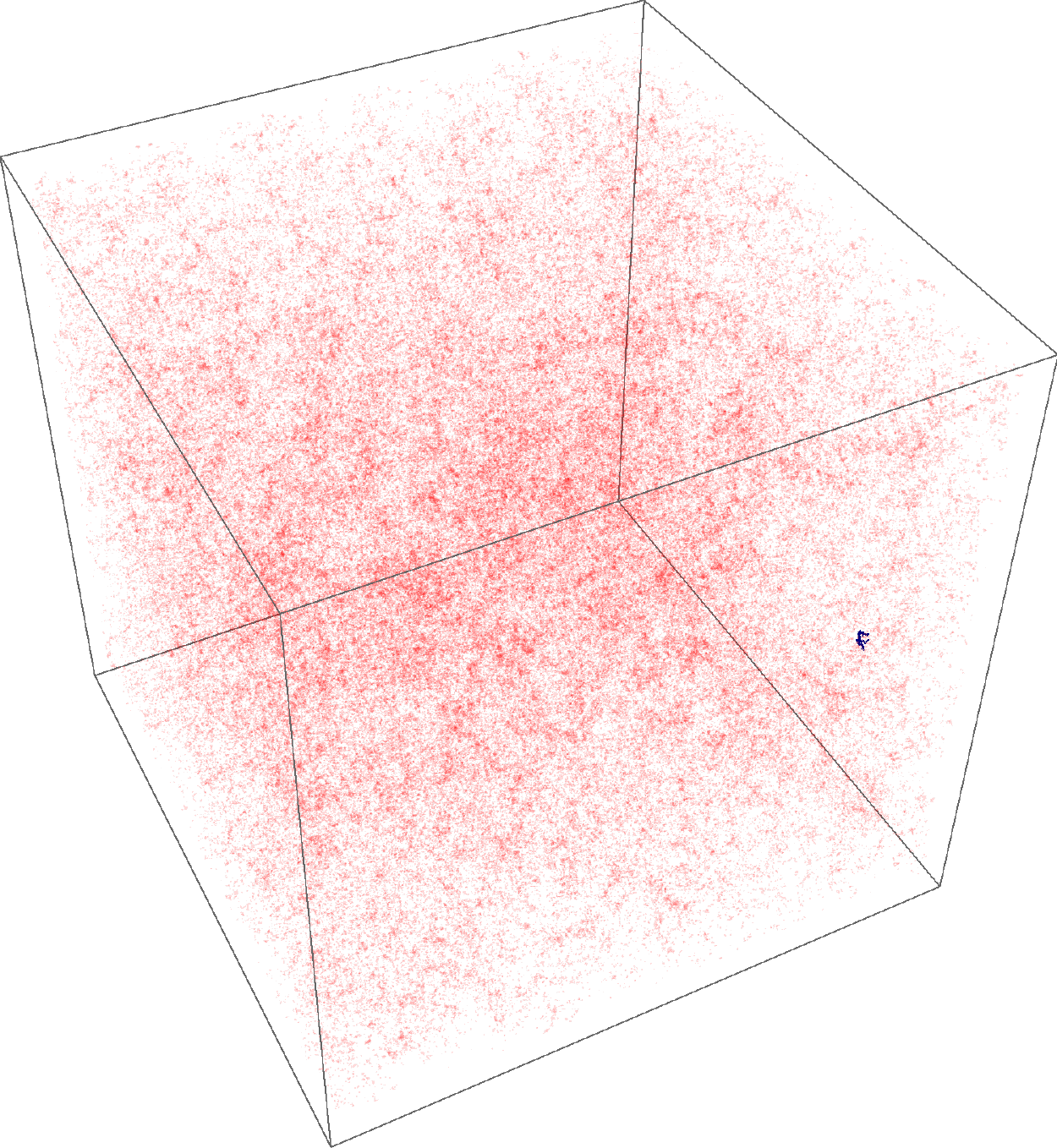}\\
\includegraphics[width=0.28\textwidth]{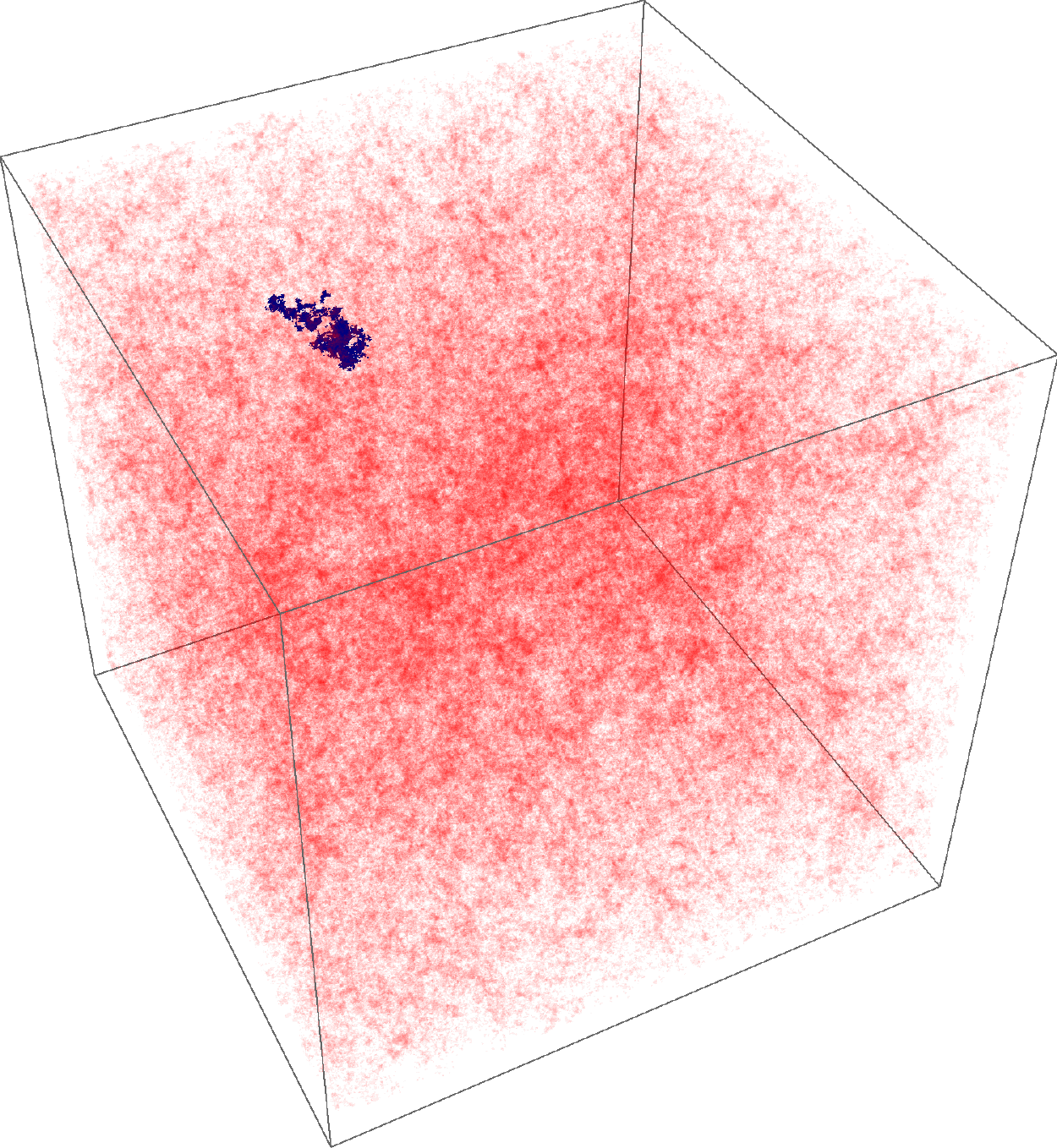}\\
\includegraphics[width=0.28\textwidth]{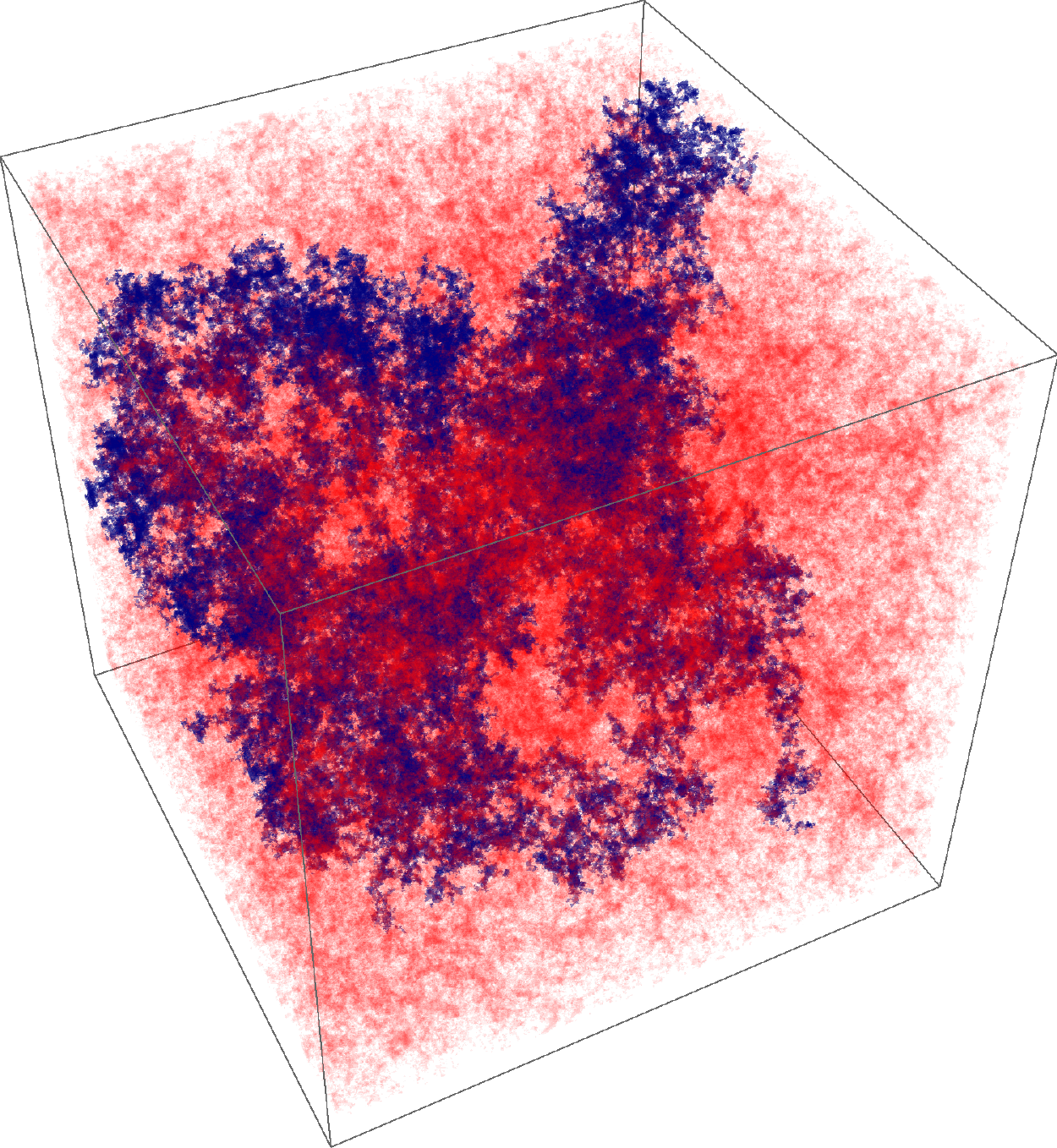}
\caption{The largest ionized region(blue) and other ionized regions(red) at $z=13.0, x_{\rm HI}=0.9925$ (top),
$z=10.75, x_{\rm HI}=0.9573$(middle), and $z=9.75, x_{\rm HI}=0.9027$ (bottom). 
The neutral regions are left transparent. } 
\label{fig3DBoxA} 
\end{figure}

The simulation box at the three redshifts is shown in Fig.~\ref{fig3DBoxA}, where the ionized regions are shown in red, 
except the largest ionized region which is colored blue, while the neutral region are left transparent. We plot the simulation 
box in this way to highlight the nature of the largest ionized region. We see that at $z=13$, there are already numerous 
ionized bubbles, but they are all very small, so that after smoothing
with a relatively large scale (8 Mpc), one would recover almost the exact Gaussian field. 
At $z=10.75$, the density of the ionized regions increased, and some of them merge to form large ionized regions, but 
even the largest ionized region is still very small. At $z=9.750$, however, even more ionized regions appear, and 
begin to get connected with each other to form a large ionized region.
This result is consistent with the findings of \cite{Furlanetto:2015hqp} that one large ionized bubble formed at around
$\bar{x}_{\rm HI}\approx 0.9$. 

\begin{figure} [!htbp]
\centering
\includegraphics[width=0.4\textwidth]{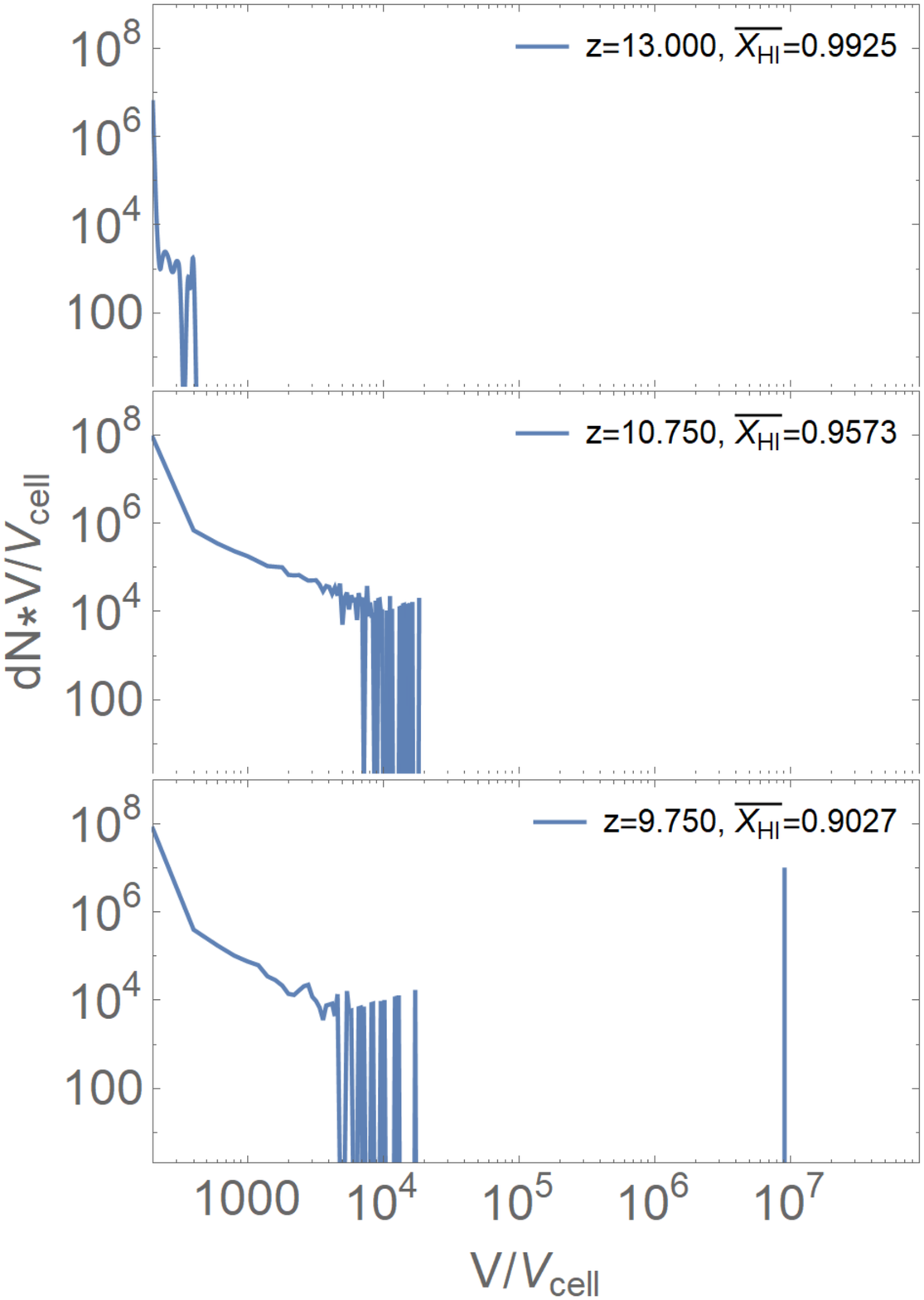} 
\caption{The size distribution of ionized regions at $z=13.000$, $\bar{x}_{\rm HI}=0.9925$ (top panel),
 $z=10.750$, $\bar{x}_{\rm HI}=0.9573$ (center panel) and $z=9.750$, $\bar{x}_{\rm HI}=0.9027$ (bottom panel).
 } 
\label{figSD1} 
\end{figure}

The size distribution of the ionized regions at each redshift is 
shown in Figure \ref{figSD1} in units of box cells. We have made the plot such that the few
 largest cells appear individually as distinct 
spikes at the large size side in the figure.  We can see that a very large ionized region appeared at $z=9.75$ box, 
its size exceeds other ionized regions by orders of magnitude.  Such a large ionized region starts to ``suppress'' 
the growth of more ionized regions, in the sense that the newly formed ionized regions are more likely to encounter
this giant one and become annexed before growing to large scale by itself. 
When the Universe is only $5\%$ ionized, there are still numerous ionized regions 
with comparable sizes of the order $10^4 ~ V_{\rm cell}$. However, the number of these is reduced when the largest ionized 
region formed at $x_{\rm HI}\sim0.9$, and the volume of other ionized regions stay much lower. The value of the
distribution function at small volume is still rather large, the small bubbles are not significantly affected. Recall that due to 
the well known biased formation of structures, the larger bubbles are likely formed in large regions with high average 
density, while the many smaller bubbles are more likely to form in regions of lower average densities,
so the isolated bubble assumption could still be applied to these after this stage.

\subsection{The Ionized Fibers stage}

The second column of Fig.~\ref{figMF} from left shows the MFs of the second stage of the reionization process. 
The three curves are for $z=8.875, x_{\rm HI}=0.792$ (red solid line),  $z=8.5, x_{\rm HI} =0.7230$(blue dashed line) and 
$z=8.25, x_{\rm HI}=0.6569$ (brown dash-dotted line). The MFs around the ionization boundary $T_b=0$ are clearly different 
from the Gaussian field in this stage.

As noted above, at $x_{\rm HI}\sim 0.9$ a large ionized region formed, which goes through the 
whole simulation box. The formation of a connected ionized region which is infinitely large, or going through the 
whole box volume in a simulation, marks the point of percolation of ionized regions \footnote{Note that here 
we distinguish percolation of ionized regions and the overlap of such regions. The former
refers to the point of forming an infinitely large ionized region (in a simulation, one which goes through the whole simulation 
box), while the latter refers that most of the ionized regions get connected with each other.  We did not distinguish 
these two but lumped them together as a single stage in \citet{Xu:2013npa}. 
As there are substantial difference in the  neutral fractions and 
redshifts of these two events, we now distinguish them. In \citet{Furlanetto:2015hqp},  ``percolation" 
refers to what we called bubble overlap here. However, the present usage is more in line with the usual 
definition of percolation theory \citep{2015PhR...578....1S}. 
So the percolation threshold is $p_c \approx 0.9$ for our simulation}. 
Not all of the ionized regions are connected at this point, however, as is clearly shown in Fig.~\ref{fig3DBoxA}. 
But as a matter of fact, even to the very late stage of reionization, there may still be 
some isolated ionized regions referred to as ``bubbles-in-island" in \citet{Xu:2013npa} 
which are not connected. At best, one can say that most of the ionized regions get connected in this era.

After the formation of this large ionized region, as redshift decreases, more and more ionized regions get connected. 
The MFs near $T_{\rm b}=0$ changed appreciably. Now for $V_0$ the drop above $T_b=0$ is more 
apparent, and it rejoins the near-Gaussian case at $V_0 \sim x_{\rm HI}$ as we discussed earlier.  
The surface area $V_1$ at $T_{\rm b}=0$ contours increase continuously, as ionized regions of
large size formed.  An additional local peak near $T_b \sim 0$ appeared
when the global neutral fraction reaches $x_{\rm HI}\sim0.70$. Meanwhile, the trough in the mean curvature $V_2$ 
and the peak in the Euler characteristic $V_3$ shifts toward left, also reaching $T_b \sim 0$ at $x_{\rm HI}\sim0.70$.
The position of these peaks and trough are slightly above $T_b = 0$ thanks to the smoothing effect.
These features in $V_1, V_2$ and $V_3$  near $T_b = 0$ indicates that there is a very large area and complicated
topology and morphology for the ionized region, which is an nature outcome of the connection of the ionized regions.
We shall refer to this point as the overlap of ionized regions.

\begin{figure} [!htbp]
\centering
\includegraphics[width=0.4\textwidth]{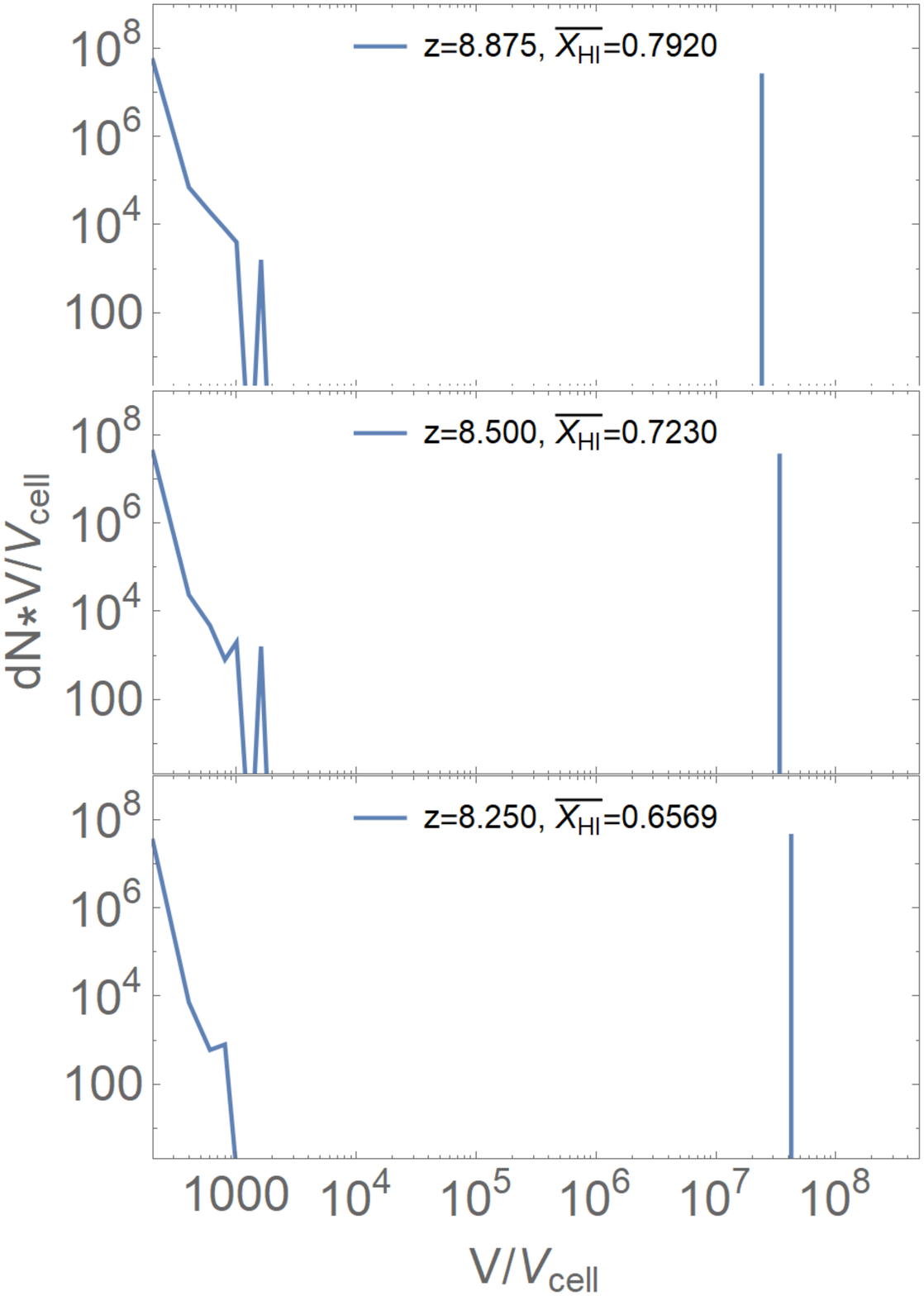} 
\caption{The size distribution of the ionized regions at $z=9.75, x_{\rm HI}=0.90275$ (top), 
$z=8.5, x_{\rm HI}=0.723$ (middle) and $z=8.25, x_{\rm HI}=0.6569$ (bottom). 
} 
\label{figSD2} 
\end{figure}

As discussed in last subsection, the largest region which now permeates much of the volume has the 
effect of annexing the other large ionized regions. The  size distribution of ionized regions is shown in Fig.\ref{figSD2}. 
The few largest ionized regions of $\sim 10^4 ~V_{\rm cell}$ in Fig.~\ref{figSD1} now disappears, 
as most of the ionized regions get connected with each other. However, the number of small bubbles remain numerous, 
showing that not all ionized regions are connected.  

\begin{figure} [!htbp]
\centering
\includegraphics[width=0.28\textwidth]{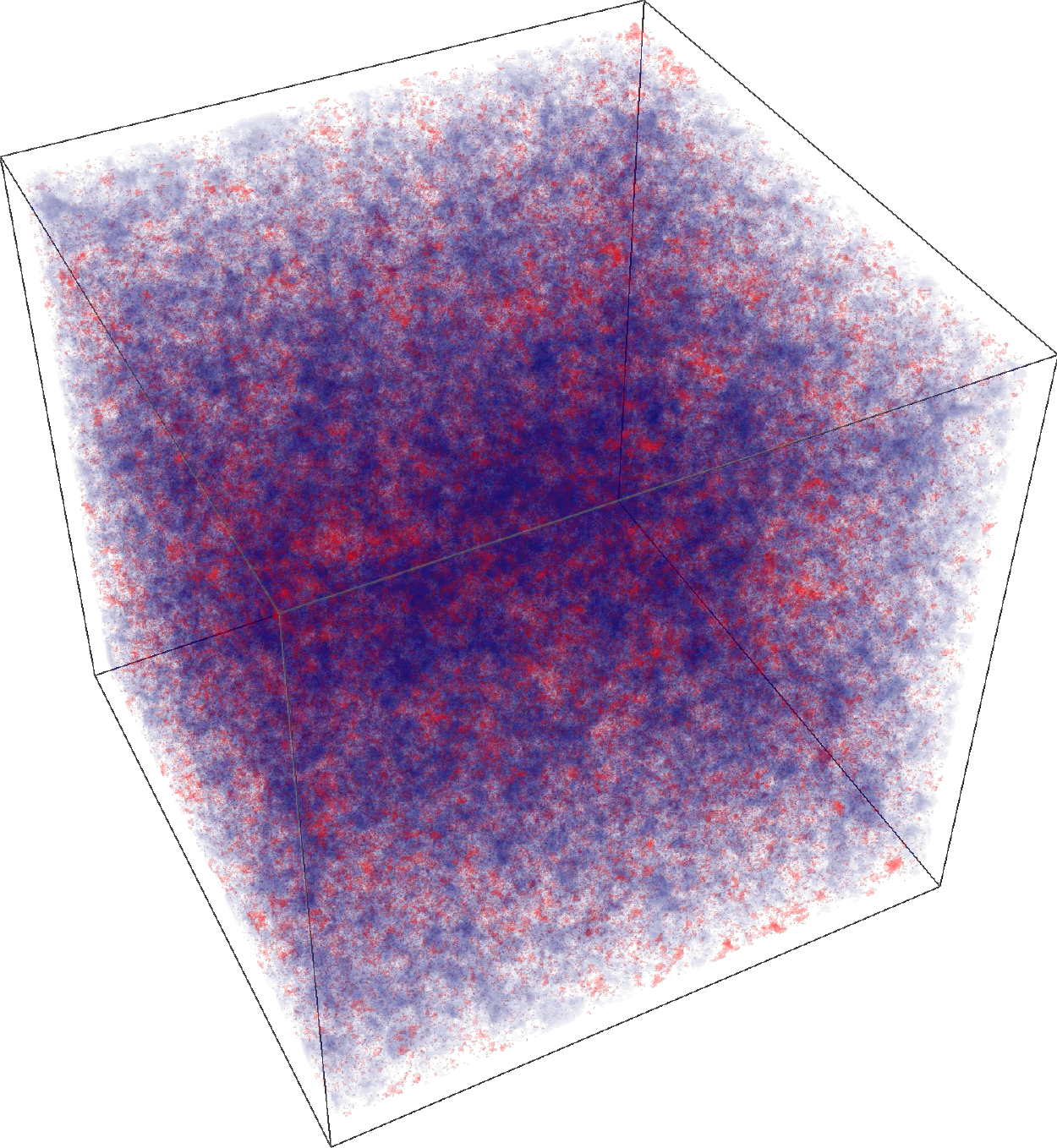} 
\caption{The largest connected ionized region at $z=8.25$.} 
\label{fig3DBoxB} 
\end{figure}

We show the largest ionized region after the overlap of ionized regions
in Figure \ref{fig3DBoxB}. It is difficult to identify the beginning of overlap visually, but the 
additional peak in the Minkowski Functional $V_1$ curve at around $T_b = 0$ give a cue for 
its occurrence. The typical global neutral fraction at which overlap happens is found to be $\bar{x}_{\rm HI}\sim0.70$, which is
consistent with the findings of \citet{Furlanetto:2015hqp} and \citet{Bag:2018zon}.

\subsection{The Sponge Stage}
The MFs during next stage of reionization is shown in the third column from left in Fig.~\ref{figMF}. Here all three 
curves are from an epoch after the overlap,  for $z= 7.875, x_{\rm HI}=0.525$ (red solid line), 
$z= 7.5, x_{\rm HI}=0.3352$ (blue dashed line), and $z=7.375, x_{\rm HI}=0.2548$.

\begin{figure}[!htbp]
\centering
\includegraphics[width=0.28\textwidth]{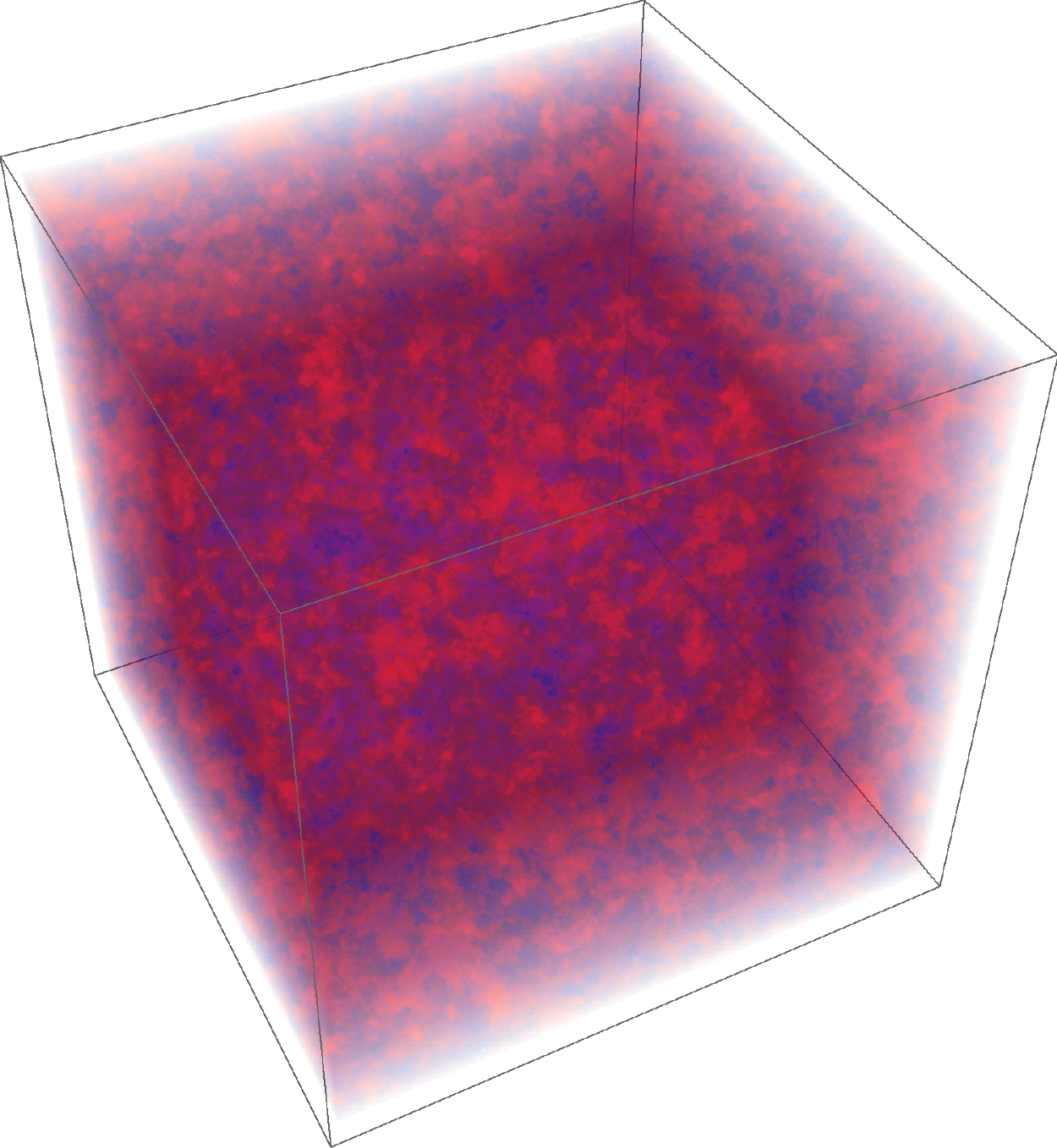}
\caption{The largest ionized region(red) and the largest neutral region(blue) at $z=7.875$. }
\label{figHBoxA}
\end{figure}

After the point of overlap, the whole box consists one large ionized region with small neutral 
regions inside, and one large neutral region, with small ionized regions inside. 
The two  are intertwining with each other in a sponge-like morphology, as shown in Figure \ref{figHBoxA}. 

The evolution of the topological structure is best shown with the 
 Euler characteristics $V_{3}$. The Euler characteristics can tell the difference between 
 a solid sphere in an empty space, which has a positive $\chi$, and holes inside a solid object, 
which has a negative $\chi$. The Euler characteristics is additive, so whether the $V_{3}$ value is positive or negative
can be used to determine whether the neutral region or ionized region is the dominating one.

\begin{figure} [!htbp]
\centering
\includegraphics[width=0.4\textwidth]{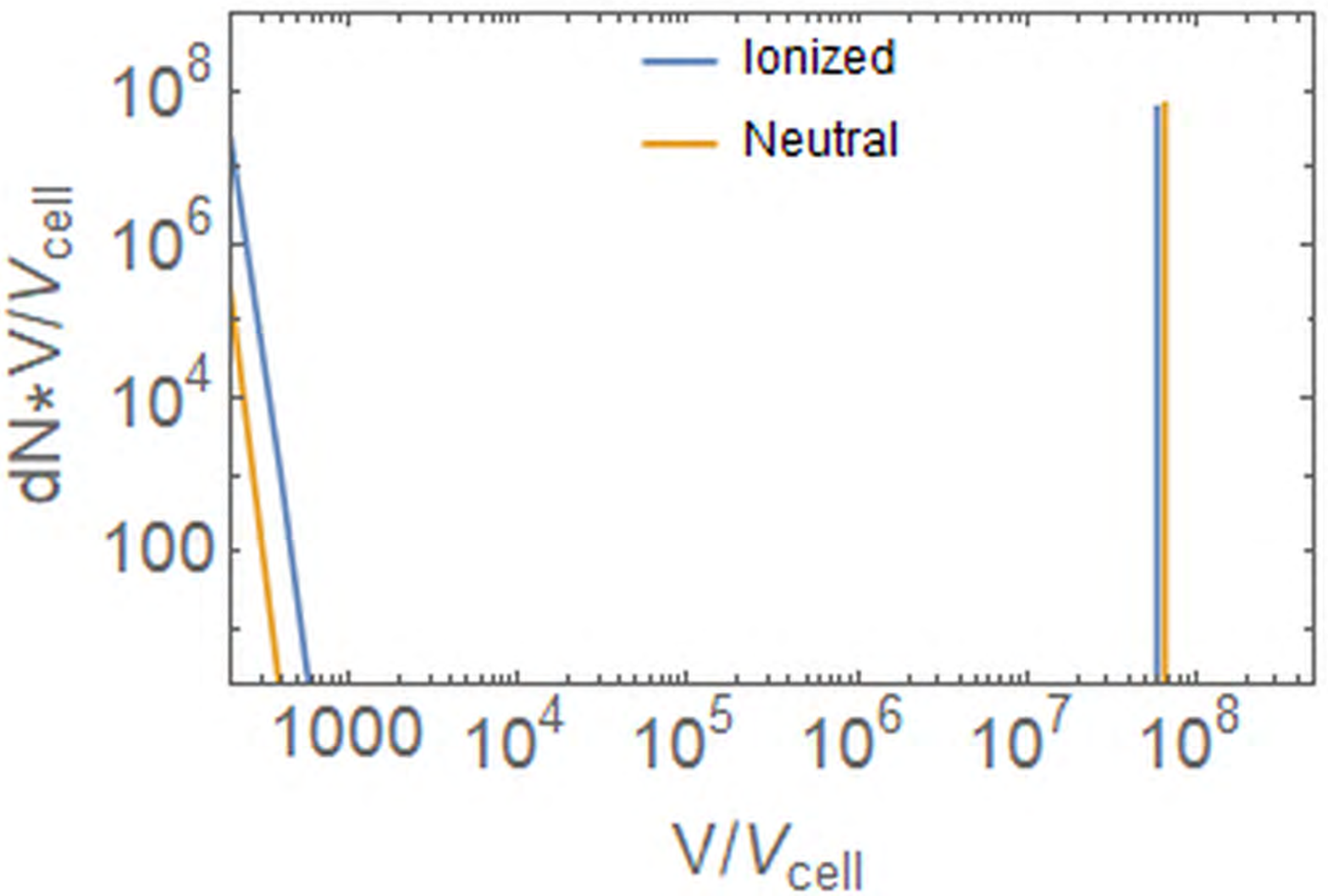} 
\caption{The size distribution of ionized regions and neutral regions at $\bar{x}_{\rm HI}=0.5250$.
} 
\label{figSD3} 
\end{figure}

Initially, $V_{3}$ has a positive peak around $T_b = 0$, reflecting the fact that even down 
to $x_{\rm HI} \sim 0.5$, the ionized regions are still in some sense surrounded by neutral regions.
When the neutral fraction drops to around 0.3, however, the $V_{3}$ at $T_b = 0$ become negative.
This is the sign that there are similar number of ``holes" (neutral regions inside ionized region) to 
that of ``balls" (ionized regions inside neutral region).  
We note that there is an asymmetry in the evolution of topology as a function of the global neutral fraction, which may 
help to explain why the analytical ``bubble model'' and semi-numerical models based on the same idea works fairly well statistically 
throughout most of the epoch of reionization.

The other MFs also show interesting evolution during this stage. The volume fraction $V_0$ now exhibits a cliff-like initial drop at 
$T_b \sim 0$, then it changes to a shallower decrease. This very steep initial drop does not reach $x_{\rm HI}$ however, 
but only down some point higher than it. In earlier discussions we noted that the smoothing  effect make the drop of $V_0$
less steep. The appearance of the very steep initial drop shows that at this stage of reionization, some very large ionized regions
are present, so that they are not affected by the smoothing. However, some of the ionized regions are still small which 
are subject to the smoothing effect, so the $V_0$ changes to a more gentle decline before reaching $x_{\rm HI}$.
Looking more carefully, there is a small bounce at the bottom of its initial drop. As $V_0(x) $ is the fraction of volume
with $T_b>x$, it should be monotonically decrease.
The peak in the surface area $V_1$ and the trough in the mean curvature $V_2$ at $T_b\sim 0$ becomes more prominent, 
consistent with the impression that the interface between the neutral and ionized regions are highly complicated for the 
sponge-like topology in this stage. 

The fact that the neutral region still surrounds the ionized regions after the Universe is $50\%$ ionized can also be seen 
using the size distribution of ionized region as shown in Figure \ref{figSD3}. Both the largest neutral region and the largest
ionized region appear as a dominant cell cluster in this figure, and they have comparable volumes.
However, the number of small ionized regions and neutral regions have a significant difference, with many more (two orders of 
magnitude higher) ionized regions than its neutral counterparts. The bubbles are much more numerous than the 
islands, hence the positive Euler characteristics.

\subsection{The Neutral Fibers Stage}

As the redshift decreases further, most of the cosmic volume become occupied by the ionized regions, 
and we expect to enter a neutral fibers stage,  which in some sense can be regarded as 
the mirror of the ionized fiber state: the neutral regions
are now embedded within the ionized regions, but most of the neutral regions are connected. 

\begin{figure} [!htbp]
\centering
\includegraphics[width=0.28\textwidth]{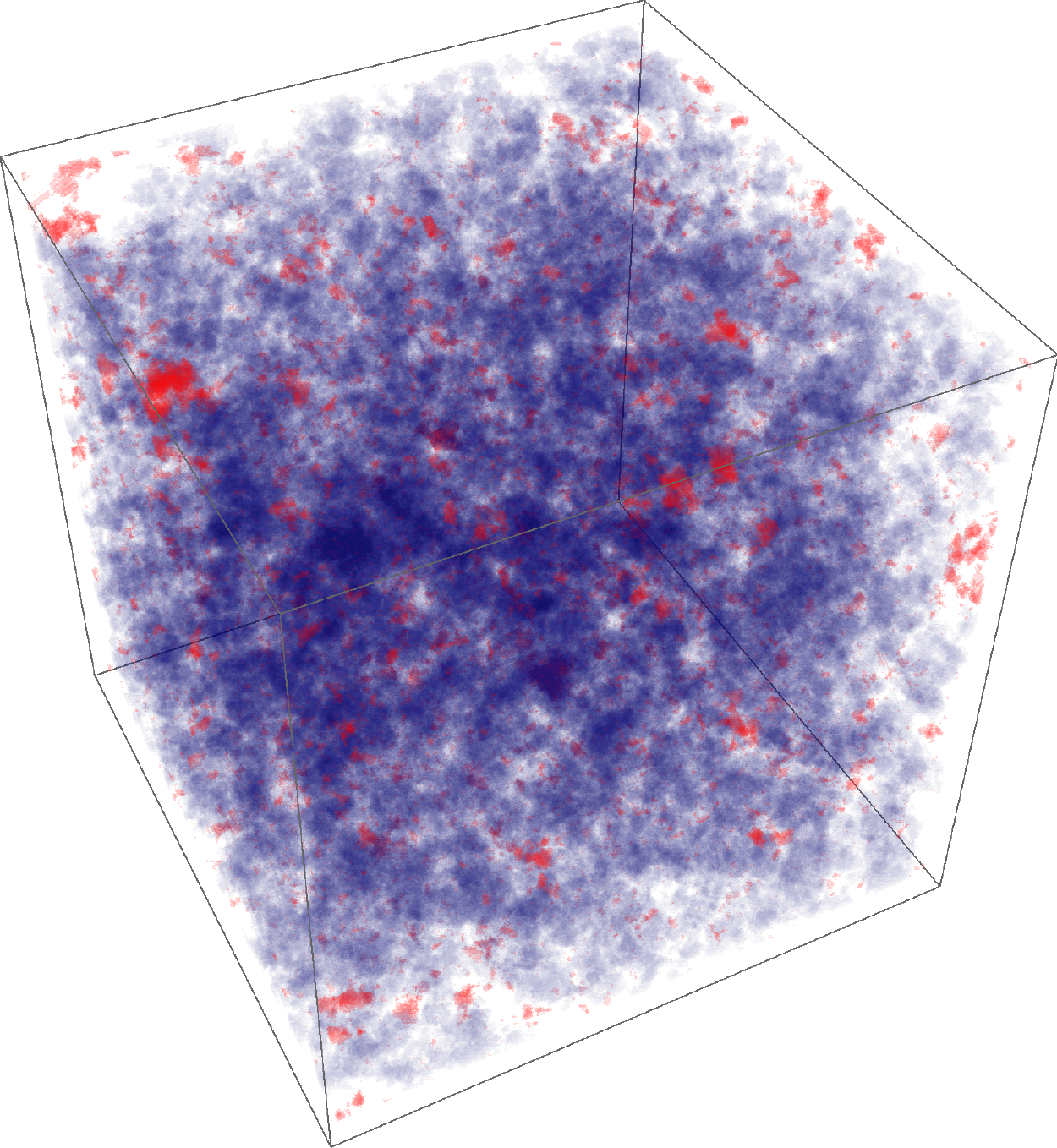} 
\includegraphics[width=0.28\textwidth]{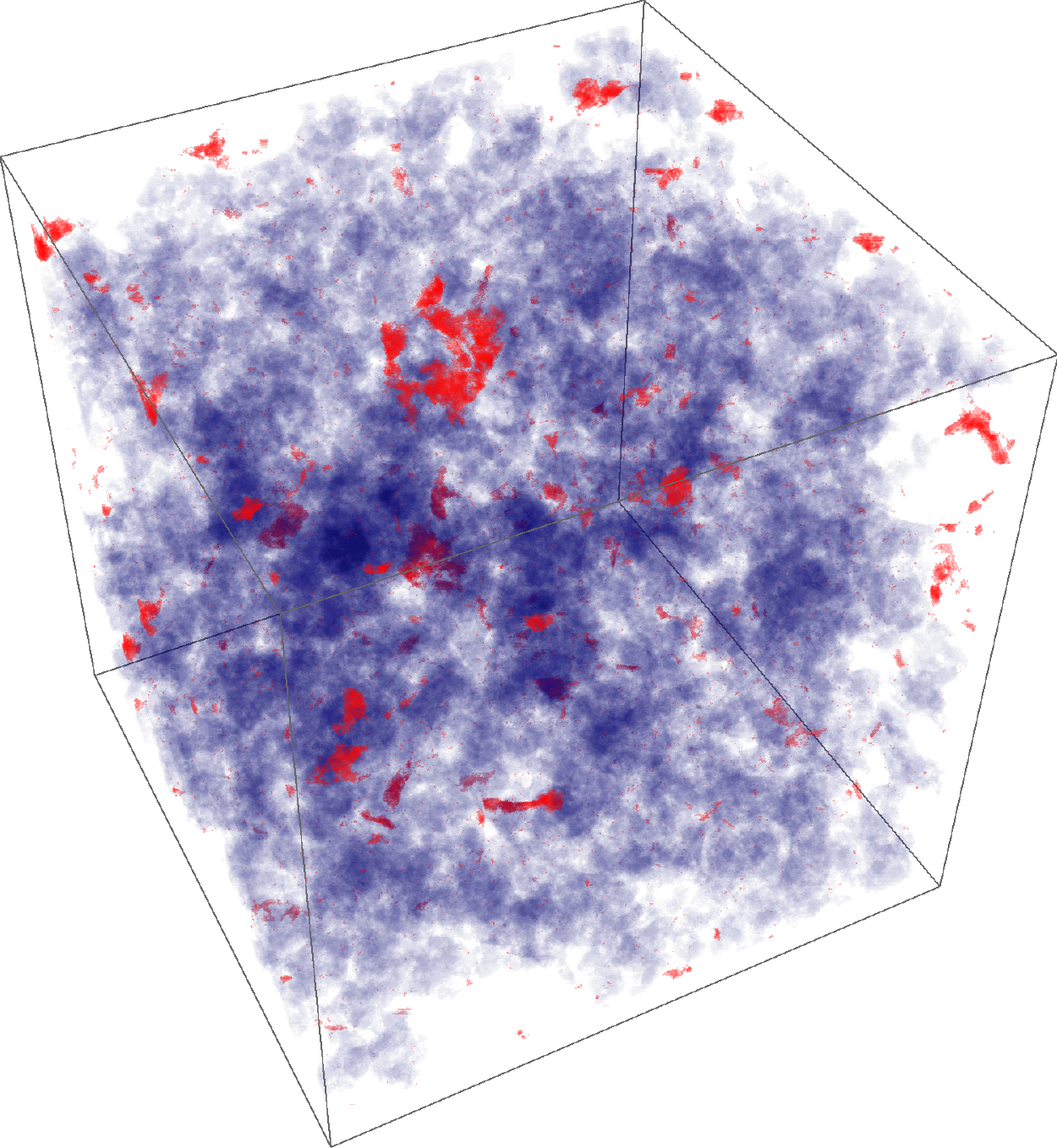} 
\caption{The largest neutral region(blue) and other neutral regions(red) at $z=7.35, x_{\rm HI}=0.2548$ (top) 
and $z=7.25, x_{\rm HI}=0.1601$. The ionized regions are left transparent.} 
\label{figNuFibersBox} 
\end{figure}

Two curves in the last column of Fig.~\ref{figMF} are the MF during this stage. The first curve, which is the same as the 
last curve of the last stage, with $z=7.375, x_{\rm HI}=0.2548$ (red solid line), and the second has 
$z=7.25, x_{\rm HI}=0.1601$ (blue dash line). 

Now the $V_0$ has an even larger steep initial drop at $T_b=0$, 
though still not down to $x_{\rm HI}$ during this steep drop due to the smoothing effect.
The surface area $V_1$ at $T_b=0$ decreases a bit as the neutral region begin to shrink.  The negative trough
of $V_2$ also become less deep. The trough of $V_3$ deepens however, showing that there are more ``tunnels"
than the "holes", and the neutral region remain to have a highly complicated topology. 
However, at $x_{\rm HI}=0.16$ the $V_{3}$ drops to a minimum, approaching the percolation threshold. 

In Fig.~\ref{figNuFibersBox} we show the largest neutral region in blue and all other neutral regions in red, while ionized 
regions are transparent. We can see the largest neutral region has a fiber-like morphology, similar to the 
ionized regions during the ionized fiber stage. At the lower redshift with $x_{\rm HI}=0.16$, 
the size of the largest fiber decrease, and some of its  branches are cut off by ionization and become small, 
independent neutral regions.

\subsection{The Neutral Islands Stage}

As the reionization progress further, the connected network of neutral regions gradually break into pieces and become 
isolated neutral islands. The last curve in the last column of Fig.~\ref{figMF}, $z=7.1, x_{\rm HI}=0.023$  shows this final stage 
of reionization.

\begin{figure}[!hbt]
\centering
\includegraphics[width=0.28\textwidth]{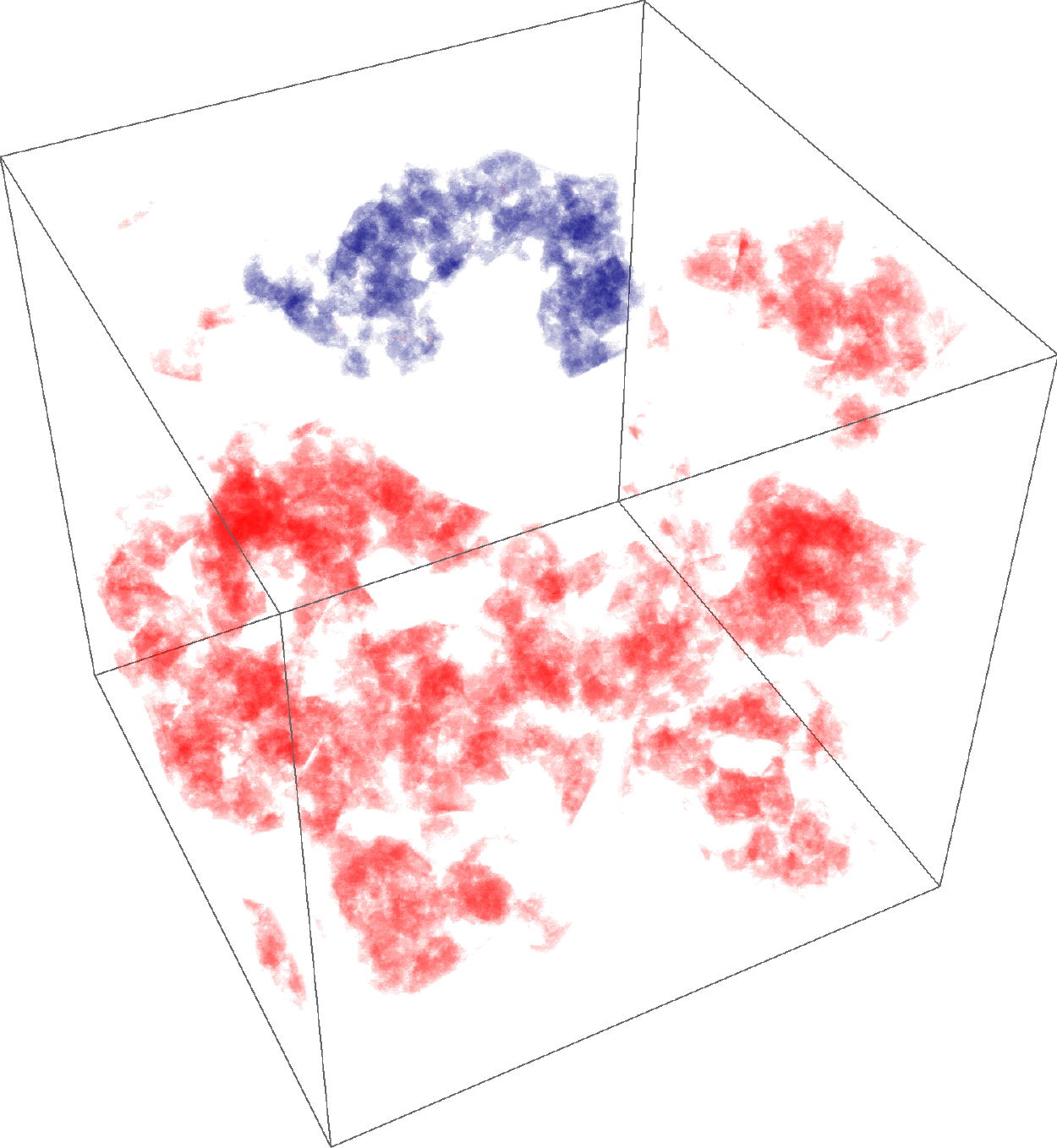} 
\caption{The largest neutral region(blue) with other neutral regions(red) as simulated with 21cmFAST 
at $z=7.1, x_{\rm HI}=0.023$, ionized cells are 
left transparent.} 
\label{figIslandBox} 
\end{figure}

 The volume fraction $V_0$ now drops as a cliff at $T_b>0$, to almost zero, then bounces back to a value 
 slightly larger than $x_{\rm HI}$,  due to smoothing.  For the surface area $V_1$, the peak at $T_b=0$ lowers as the 
 number and area of the neutral region is decreasing, and at higher $T_b$ it almost vanished. The trough at 
 $T_b=0$ for the mean curvature $V_2$ and the Euler characteristic $V_3$ also has the same vanishing trend. Furthermore, 
 during the Neutral Fibers stage there is still some peaks in $T_b \gg 0$, indicating the presence of some dense clumps 
 inside the large neutral region, these now also disappeared. At this point, most neutral regions are 
isolated, contributing the number of ``holes'', and the island model description becomes applicable.

After the neutral islands become isolated, they continue to shrink and disappear, resulting the decreasing
value of $V_3$ around $T_b \sim 0$. Eventually, all the islands are ionized and $V_3$ 
returns to 0 as the reionization is completed. In Fig.\ref{figIslandBox} we show the largest neutral region 
in blue, and other neutral regions in red. Here we see even the largest neutral region is now relatively small, but there
are still numerous small islands. However, all of these neutral will be reionized eventually.

\begin{figure} [!htbp]
\centering
\includegraphics[width=0.35\textwidth]{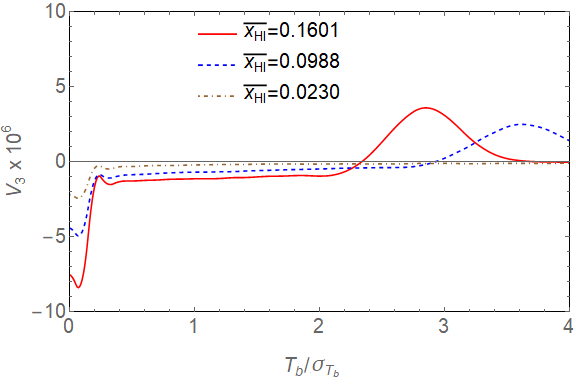} \\
\includegraphics[width=0.35\textwidth]{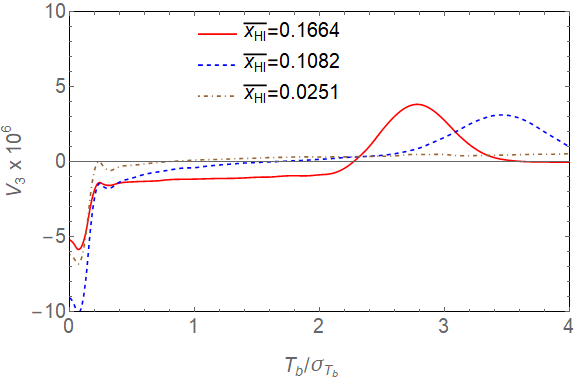} 
\caption{The evolution of $V_{3}$ simulated with 21cmFAST ({\it top panel}) and IslandFAST ({\it bottom panel}).
} 
\label{figDiff} 
\end{figure}

\section{IslandFAST vs. 21cmFAST}

\begin{figure*} [!ht]
\centering
\subfigure[$\bar{x}_{\rm HI}=0.1601$, 21cmFAST] { \label{fig:a} 
\includegraphics[width=0.28\textwidth]{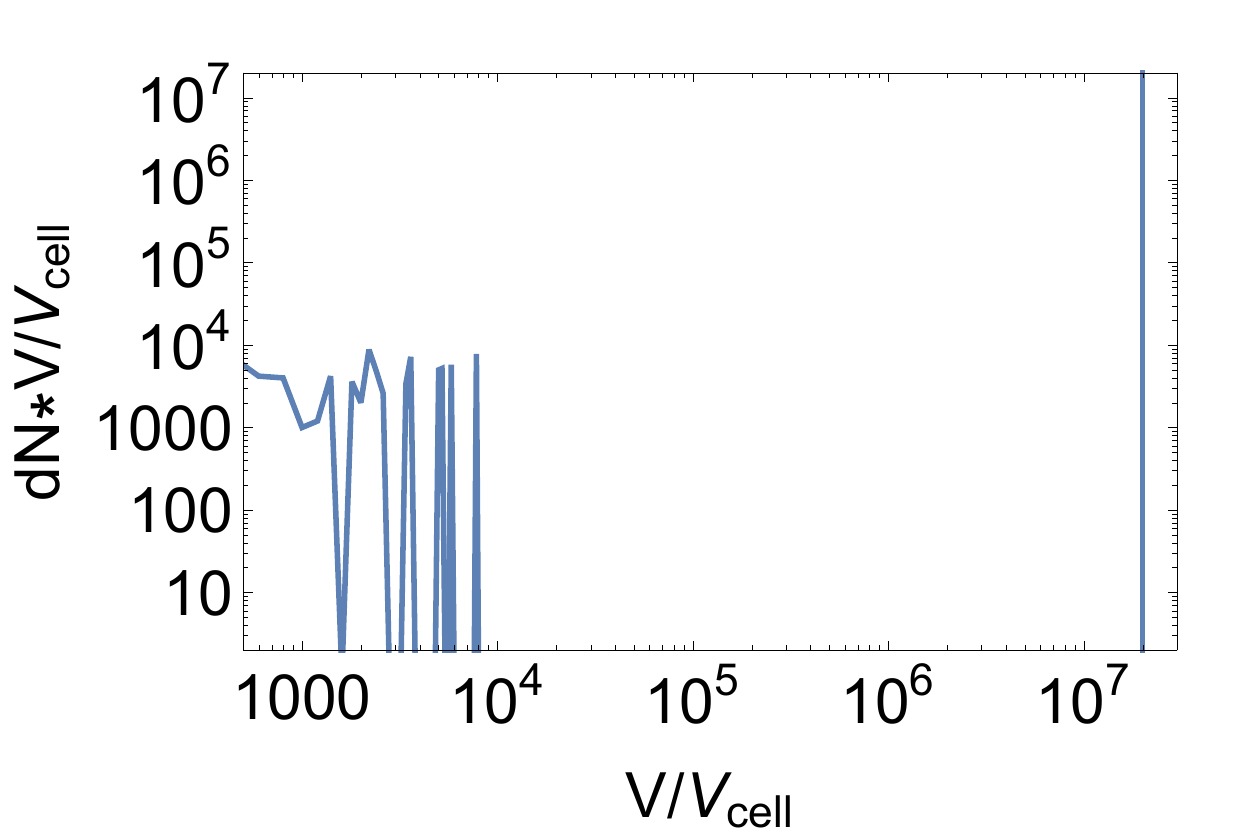} 
}  
\subfigure[$\bar{x}_{\rm HI}=0.0988$, 21cmFAST] { \label{fig:b} 
\includegraphics[width=0.28\textwidth]{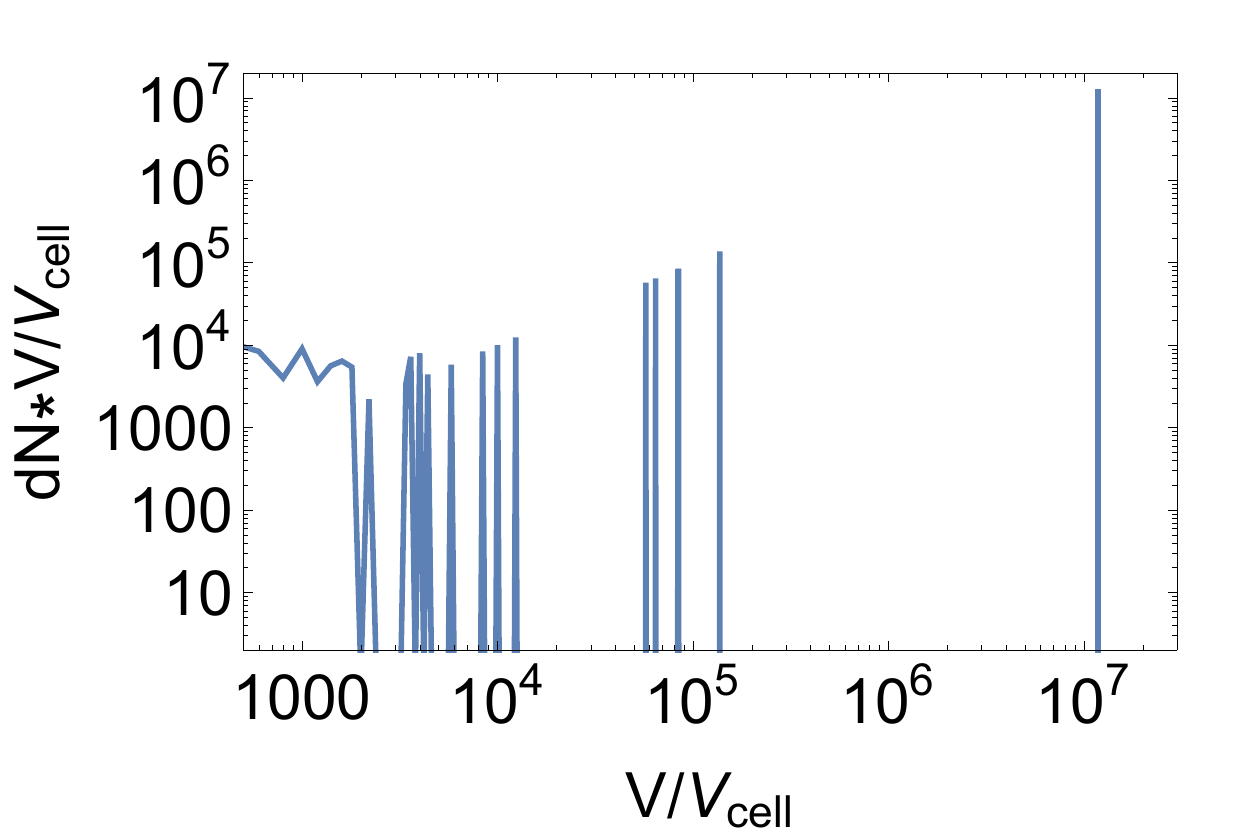} 
} 
\subfigure[$\bar{x}_{\rm HI}=0.0687$, 21cmFAST] { \label{fig:c} 
\includegraphics[width=0.28\textwidth]{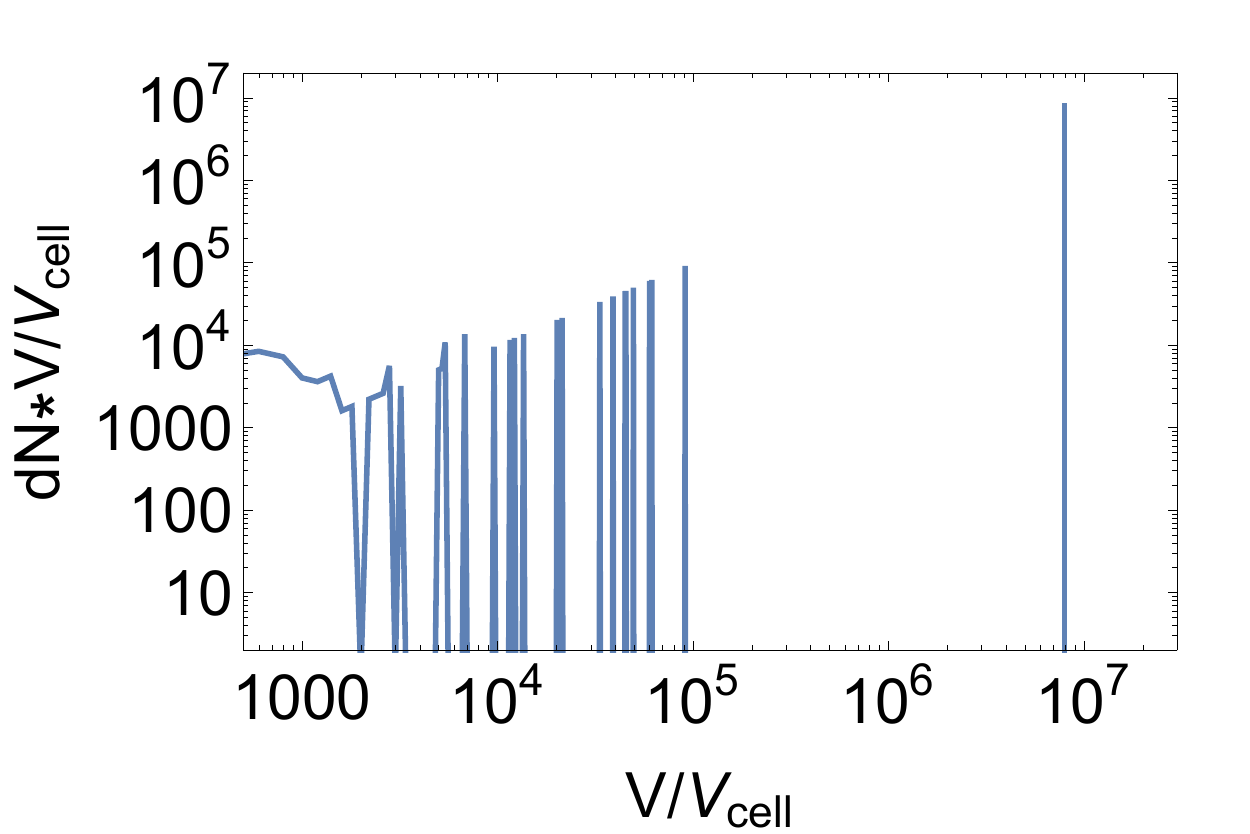} 
}   \\
\subfigure[$\bar{x}_{\rm HI}=0.1664$, IslandFAST] { \label{fig:d} 
\includegraphics[width=0.28\textwidth]{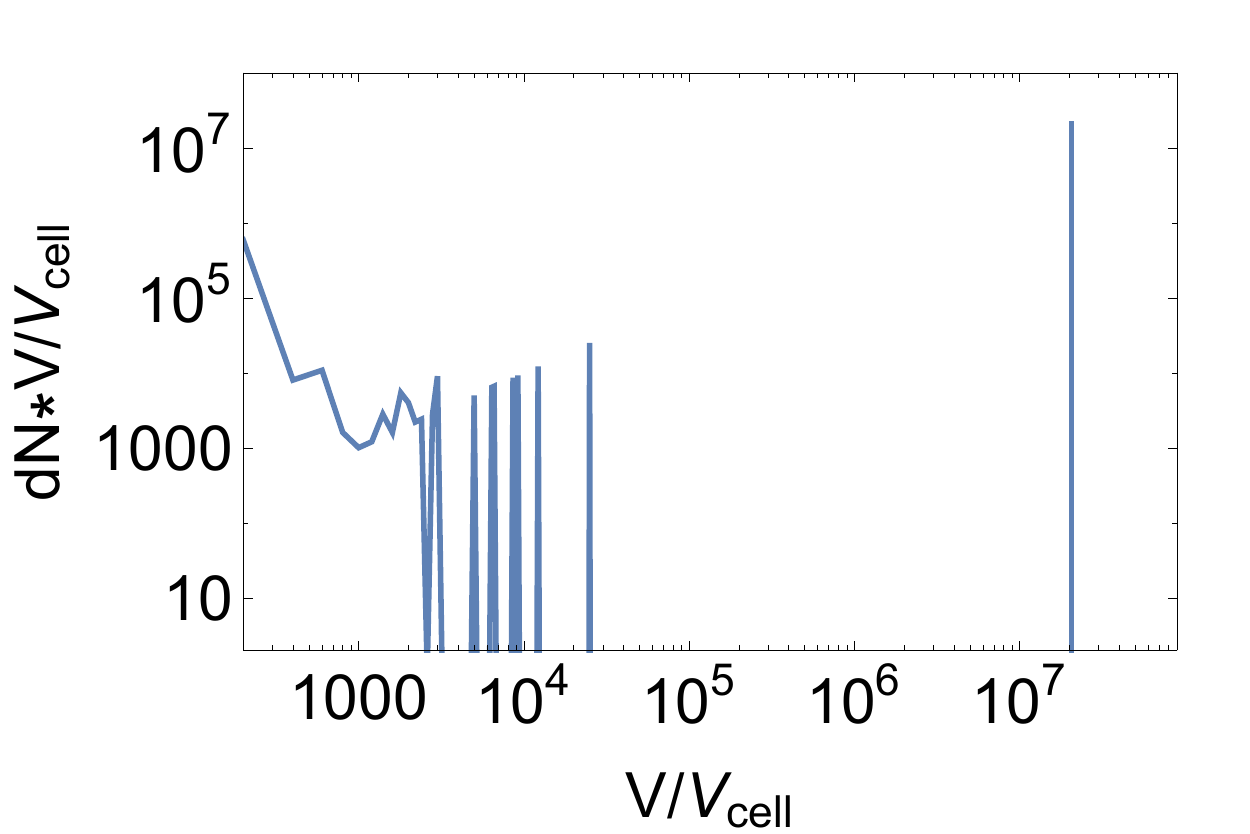} 
}  
\subfigure[$\bar{x}_{\rm HI}=0.1082$, IslandFAST] { \label{fig:e} 
\includegraphics[width=0.28\textwidth]{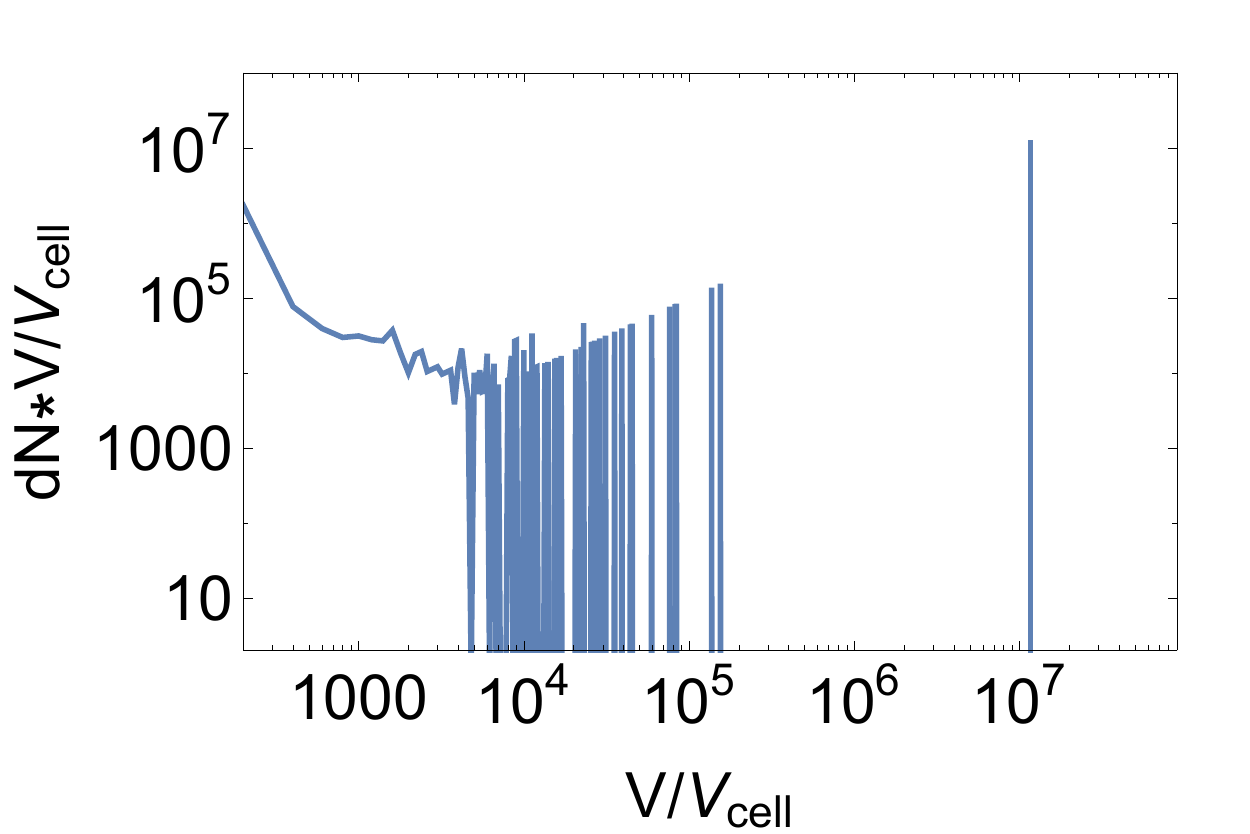} 
} 
\subfigure[$\bar{x}_{\rm HI}=0.0769$, IslandFAST] { \label{fig:f} 
\includegraphics[width=0.28\textwidth]{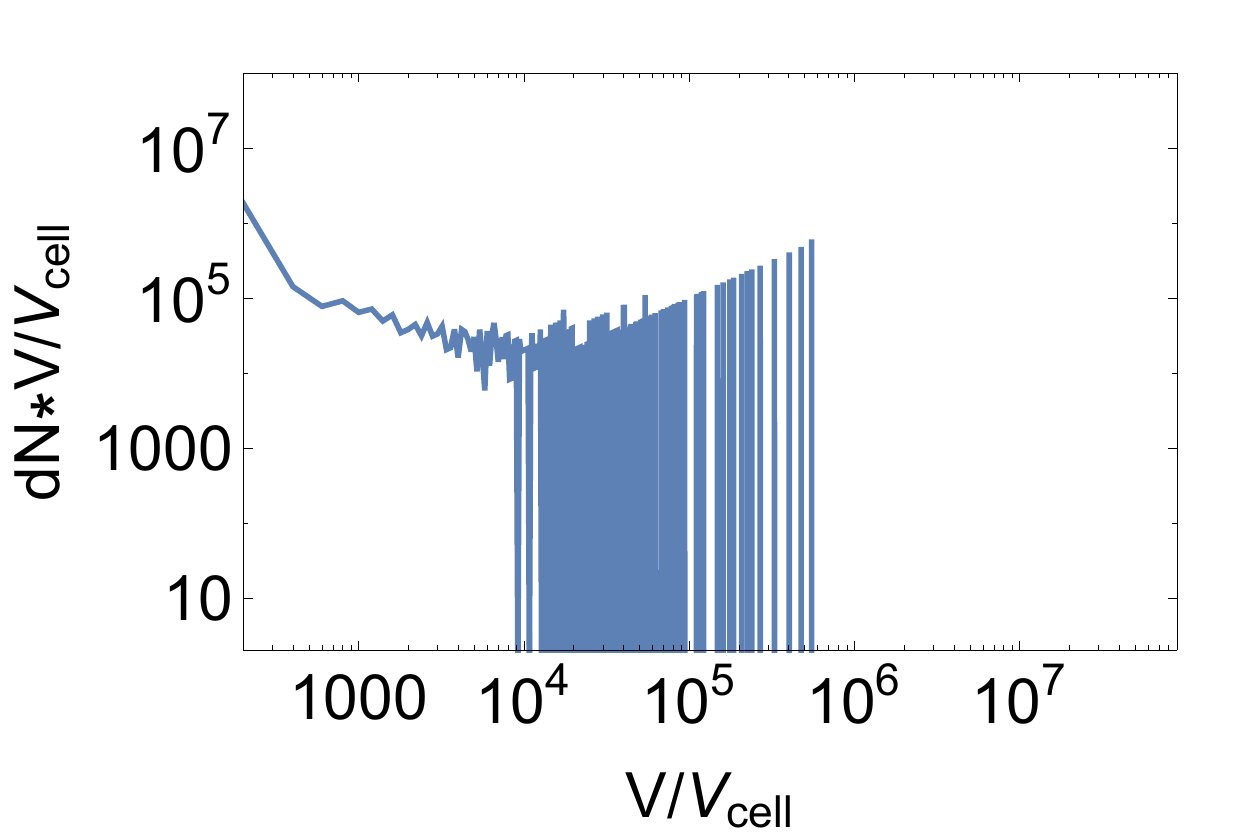} 
} 
\caption{Size distribution of neutral regions in each neutral fraction from $x_{\rm HI}\sim0.17$ to 
$\bar{x}_{\rm HI}\sim0.07$. The top panel are results of 21cmFAST, and the bottom panel are results for islandFAST.\\} 
\label{figSD4} 
\end{figure*}

In the above, the evolution of the ionization field were obtained by using the 21cmFAST  semi-numerical simulation.
However, the external photons from the ionizing background may play an important role in the later stages of reionization. 
This is taken into account by the IslandFAST model. Here we make a re-analysis of the later stages by comparing the 
results of the 21cmFAST and IslandFAST simulations. The IslandFAST requires an input on the external ionizing 
background at the start, though afterwards the ionizing background at each later redshift is derived self-consistently. 
Here we started the IslandFAST simulation at $x_{\rm HI}=0.17$, using the 
output of the 21cmFAST at that redshift as initial condition. We then compare the subsequent topological 
evolution for the two models. 

The evolution of $V_{3}$ as simulated with IslandFAST and 21cmFAST are shown in Figure \ref{figDiff}. As the 
evolution in the two models are different, we take the results at approximately equal neutral fractions. 

For the 21 cm field simulated with 21cmFAST, the $V_{3}$ at $T_b=0$ simply increases monotonically during this era, 
indicating a continuous decrease in the number of individual islands. For the IslandFAST, however, 
$V_{3}$ at $T_b = 0$ first decrease to a lower minimum, then begin to increase.
The further decrease in $V_{3}$ for the IslandFAST simulation reflects the process that the relatively large 
neutral regions are being divided into smaller ones because of the bubbles-in-island effect, as well as the 
increasing ionizing background. This process creates more ``holes", resulting in a more negative 
$V_{3}$ around $T_b = 0$. Eventually, the isolated neutral islands are consumed by the ionizing background, 
and similar to the case with 21cmFAST, the $V_3$ at $T_b=0$ eventually vanishes.

\begin{figure}[!htbp]
\centering
\includegraphics[width=0.4\textwidth]{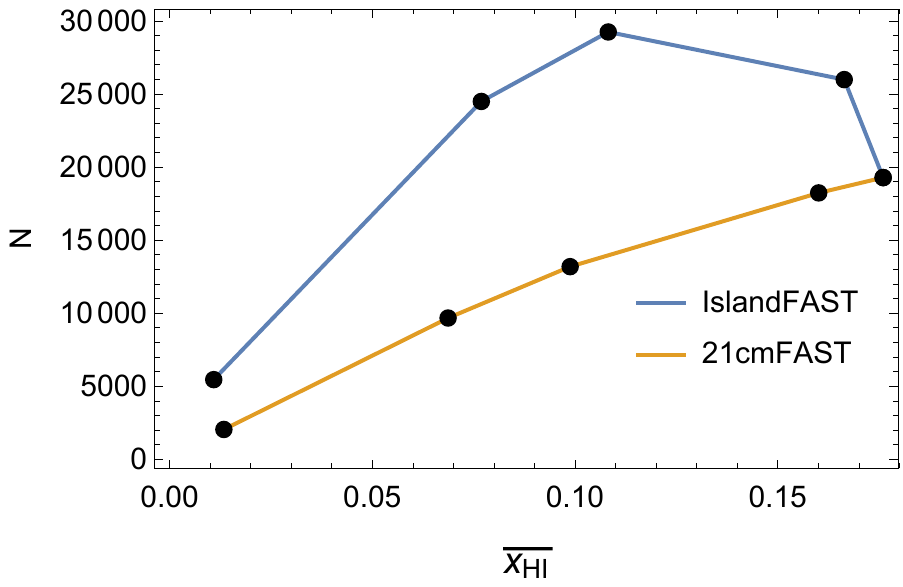}
\caption{The number of islands as a function of average neutral fraction as simulated with 21cmFAST and IslandFAST.}
\label{figIslandnumber}
\end{figure}

In Fig.~\ref{figSD4} we plot  the size distribution function of the two simulations at three comparable 
neutral fractions (the neutral fractions are derived from the simulation snapshots, so they can not be made exactly equal).
The top panels (a,b,c) show the 21cmFAST results, while the bottom panels (d,e,f) show the IslandFAST result.  
For 21cmFAST, the neutral region simply shrinks as the neutral fraction decreases, so that the whole neutral region size distribution simply shifts toward smaller ones, but a very large neutral cluster is always present. For IslandFAST, 
however, at certain point the largest neutral region breaks into disconnected pieces around $x_{\rm HI}\sim0.1$, so the 
largest island disappears, while the number of smaller ones is nearly constant or even get an increase during this 
process.  This evolution in the  number of neutral islands  as a function of average
neutral fraction is also plotted in Fig.~\ref{figIslandnumber}.  At the starting point 
$x_{\rm HI}=0.17$, the number of islands in the 21cmFAST simulation and the IslandFAST simulation are the same. 
As  $x_{\rm HI}$ decreases, however, the two models diverge. For 21cmFAST, the  number of neutral regions as well
as the volume of the largest neutral region decreases monotonically. For IslandFAST, as $x_{\rm HI}$ drops 
to $\approx 0.1$, the large neutral region breaks into a number of smaller islands, with a 
sudden increase of the number of neutral regions. Of course, for either code, the number of islands approaches zero
with $x_{\rm HI} \to 0$.

From these results, we observe that during the late stage of the reionization, the IslandFAST simulation predicts that
the largest neutral region gets cut off into multiple pieces, while in the 21cmFAST it shrinks, though at some point
it must also be broken off into two or a few pieces, as it no longer percolate through the whole box. 
This is not surprising, the IslandFAST takes the ionization caused by external photons into account explicitly, 
such photons are more uniformly distributed, and would more likely to cut through the links between the neutral regions
at the same time, and this perhaps provides a better description of the topological evolution during the late stage of EoR.

\section{Observablity}

We now study the feasibility of using the MFs to distinguish the 
various stages. The reionization process are very complicated and there are a very wide range of possibilities, it is not possible 
to cover or even explore all these possibilities here. For example, For example, if the ionizing photon spectrum are 
produced by quasars or decaying particles and very hard, many regions will be partially ionized and the scenarios 
would be entirely different. Instead, we  will use our fiducial model to illustrate how the MFs 
can be used as a quantitative tool to distinguish these stages, and similar analysis with the MFs can be performed on models
not too different. For simplicity, we select one feature for each stage transition and list them in Table \ref{table:1}, but 
we do NOT claim that these are universal criteria applicable to all models. 

\begin{table} [htbp]   
\begin{center}
\caption{A summary of the MF features at transition points}
\begin{tabular}{lll}  
\hline  \hline
Stage$^\star$  & $\bar{x}^c_{\rm HI} $ & Feature\\  
\hline  
1 $\to$ 2 & 0.9 & The fractional difference of the \\
&& two peaks of $V_3 > 0.1$   \\  
\hline
2 $\to$ 3 & 0.7 & The trough in $V_2$ shifts leftwards \\
 && to $T_b \approx 0$ \\  
\hline
3 $\to$ 4 & 0.3 & $V_3$ around $T_b\sim0$ becomes negative \\
\hline
4 $\to$ 5 & 0.1 or 0.16$^\dagger$ & The trough in $V_3$ reaches its minimum \\
\hline  
\hline
\end{tabular}
\end{center}
\footnotesize $^\star$ The numbered stages are 1.the Ionized Bubbles stage, 2. the Ionized Fibers stage, 
3. the Sponge stage, 4. the Neutral Fibers stage and 5. Neutral Islands stage.\\
\footnotesize $^\dagger$ The mean neutral fraction at the last transition is about 0.16 in 21cmFAST and about 0.10 in islandFAST.
\label{table:1}  
\end{table}  

For the Square Kilometer Array phase one low frequency experiment (SKA1-low), the 
angular and frequency resolutions are much finer than the comoving smoothing scale (16 Mpc) used for our MF measurement, 
so instrumental smearing is negligible. We assume the per pixel noise is Gaussian and given by (e.g. \citealt{Watkinson:2013fea}):
\begin{eqnarray}
\sigma_{\rm{noise}}&=&2.9 \left(\frac{10^5\rm{m^2}}{A_{\rm{eff}}}\right)\left(\frac{10'}{\Delta\theta}\right)^2 
\left(\frac{1+z}{10.0}\right)^{4.6}\nonumber \\
&&\times \sqrt{\frac{1\,\rm{MHz}}{\Delta\nu}\frac{100\, \rm{hours}}{t_{\rm{int}}}} \rm{mK},
\end{eqnarray}
with the system temperature $T_{\rm sys}\approx T_{\rm sky}$ with \citep{10.1093/mnras/stv2601}
\begin{equation}
 T_{\rm sky} = 180 \left(\frac{\nu}{180 \MHz}\right)^{-2.6} \rm{K}.
\end{equation}
The array sensitivity $A_{\rm{eff}}/T_{\rm{sys}}$ is modeled according to the SKA performance document\footnote{\url{https://astronomers.skatelescope.org/wp-content/uploads/2017/10/SKA-TEL-SKO-0000818-01_SKA1_Science_Perform.pdf}}, 
and we assume an integration time of 1000 hours. For $z=8.5$ and $\bar{x}_{\rm HI}=0.72$ in 
our fiducial model, the 
signal to noise ratio  $\approx$10 per smoothing pixel ($(16 \Mpc)^3$).

\subsection{The measurement error of MFs}
Both the 21cm temperature and the thermal noise on each pixel are random. To estimate the error on MFs, 
we generate different realizations of the 21cm field as well as the noise, and compute the sample variance. 

\begin{figure} [!htbp]
\centering
\includegraphics[width=0.4\textwidth]{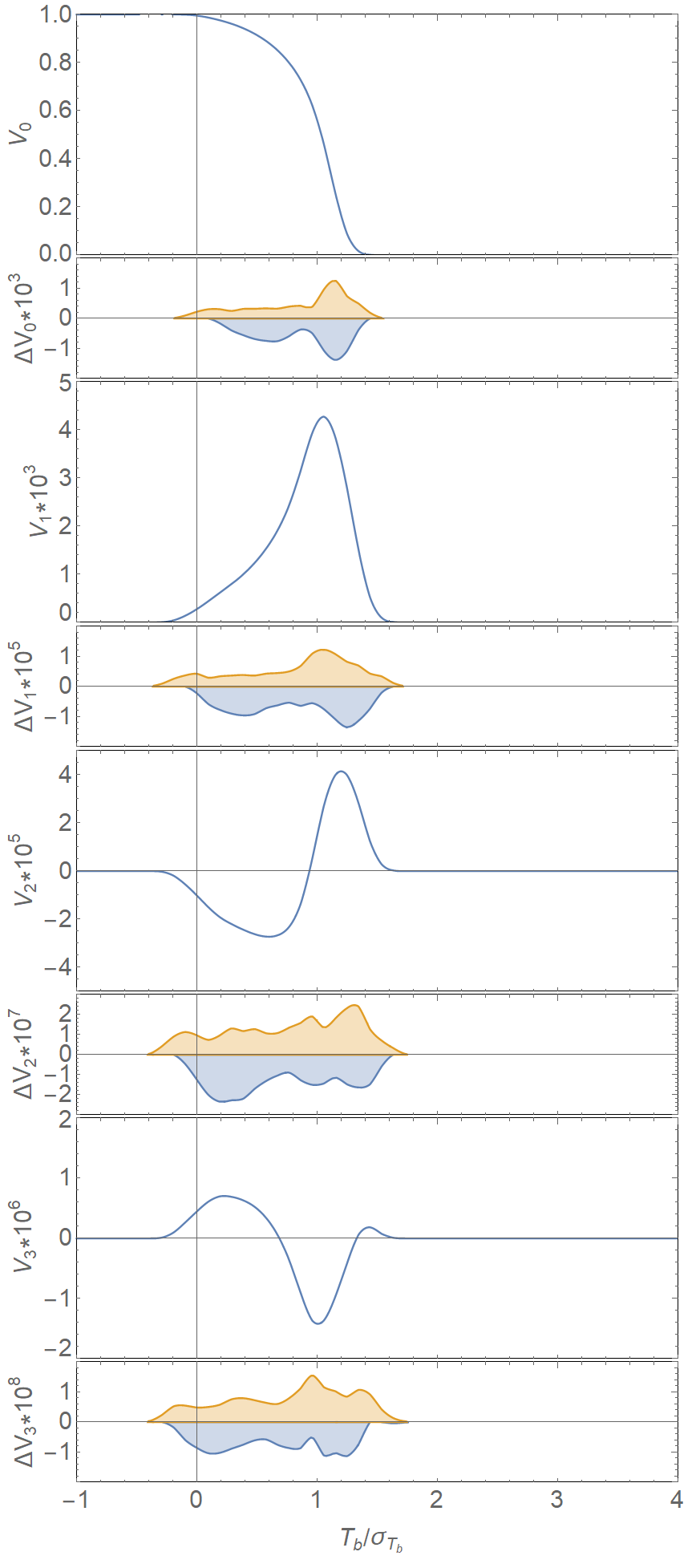}
\caption{A typical set of MFs with cosmic variance denoted as $\Delta V_{i}$ at $\bar{x}_{HI}\sim0.72$. 
The orange shadow (above the y-axis) represents the sampling error from 30 different realizations of thermal noise. 
The blue shadow (below the y-axis) represents the sampling error derived from the 8 different realizations of simulation.}
\label{MFs with noise} 
\end{figure}

First we consider the effect of cosmic variance of the 21cm field. We re-simulate the reionization process with 8 
independent realizations using the same parameters. The results are shown as  
the blue shadows (below the y-axis) in Fig.~\ref{MFs with noise}. As expected, the MFs in the different realizations
are similar and the deviations are small, showing that the MFs are indeed good statistical observables.  This is achieved with 
the 1 Gpc box, which at $z\sim 7$ corresponds to an $6.5^\circ \times 6.5^\circ$ area.  For comparison, a fiducial value of the SKA 
deep field is about  $20 \deg^2$ \citep{Greig:2019tcg}. Note that the error of the MFs approximately scales as $V^{-1/2}$, as
we verified using the simulation.

\begin{figure} [!htbp]
\centering
\includegraphics[width=0.4\textwidth]{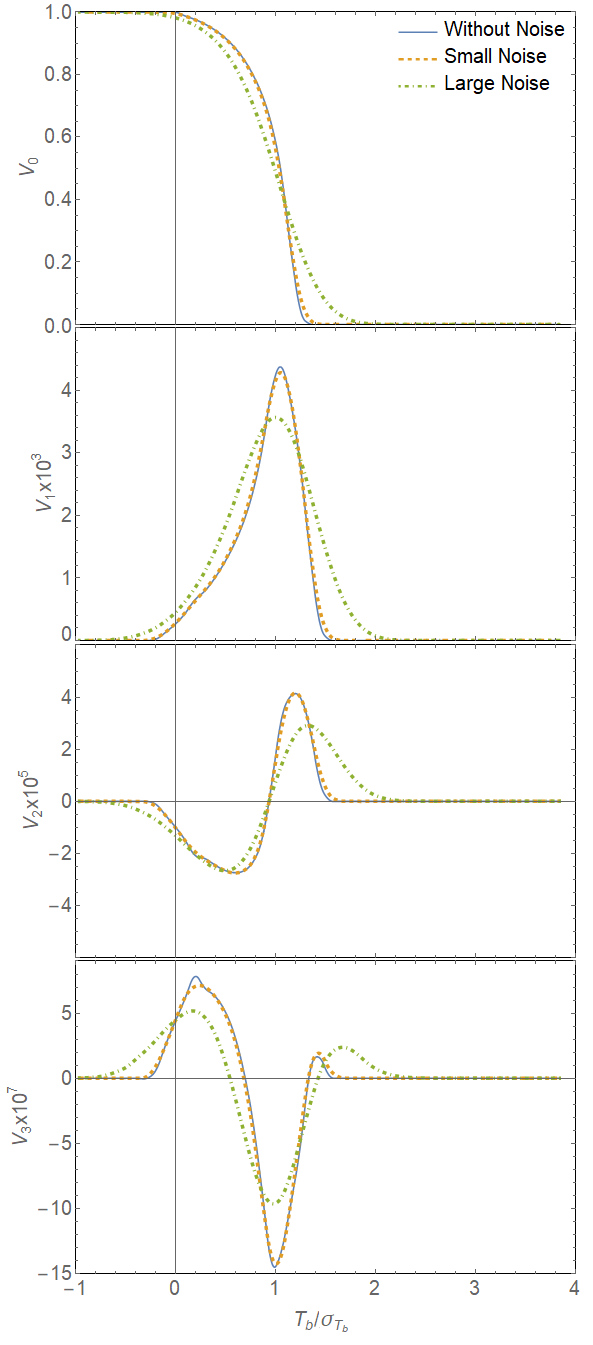}
\caption{MFs at $\bar{x}_{\rm HI}\sim0.72$ with no noise, small noise ($\rm SNR \sim 10$) and large noise
($\rm SNR \sim 3$). The $V_2$ and $V_3$ with noise are multiplied by 10 for better view.}
\label{MFs noise} 
\end{figure}

Second, we generate 30 different realizations of thermal noise, and plot the sample variance 
with the orange shadow (upper half of the y-axis) in Fig.~\ref{MFs with noise}. The variance between them are also quite small. 
However, the thermal noise has a Gaussian distribution, as a result its impacts on the Minkowski functionals are biased. 
The observed 21cm field, which is a superposition of the 21cm field and the thermal noise has a more Gaussian distribution than 
the 21cm field alone.  The features in the MFs come mainly from the topology of the $T_{\rm b}=0$ contours which 
correspond to the boundaries between neutral and ionized regions, i.e. ionizing fronts, such a contour 
is sensitive to the addition of the  thermal noise, as already pointed out in previous works on the bubble and island statistics 
\citep{Giri:2019pxr, Giri:2017nty}. In  Fig.~\ref{MFs noise}, we show the MFs for the no noise case, the small noise case
which corresponding to the one simulated above, and the large noise case which is obtained by  magnifying the thermal 
noise by a factor of 3 so that the signal to noise ratio is $\sim$ 3, the results are shown in  Fig.~\ref{MFs noise}. 
The MFs of the small noise case are almost identical to those of the no noise case. However, in the large noise case, 
 the shape of MFs becomes very similar to the MFs of Gaussian random field as illustrated in Fig.~\ref{figGauss}, effectively
 wiping out the features of the percolation cluster. The MFs are only effective reionization topology indicators when
 large signal to noise ratio observation data is available.

\begin{figure*} [!htbp]
\centering
\includegraphics[width=0.9\textwidth]{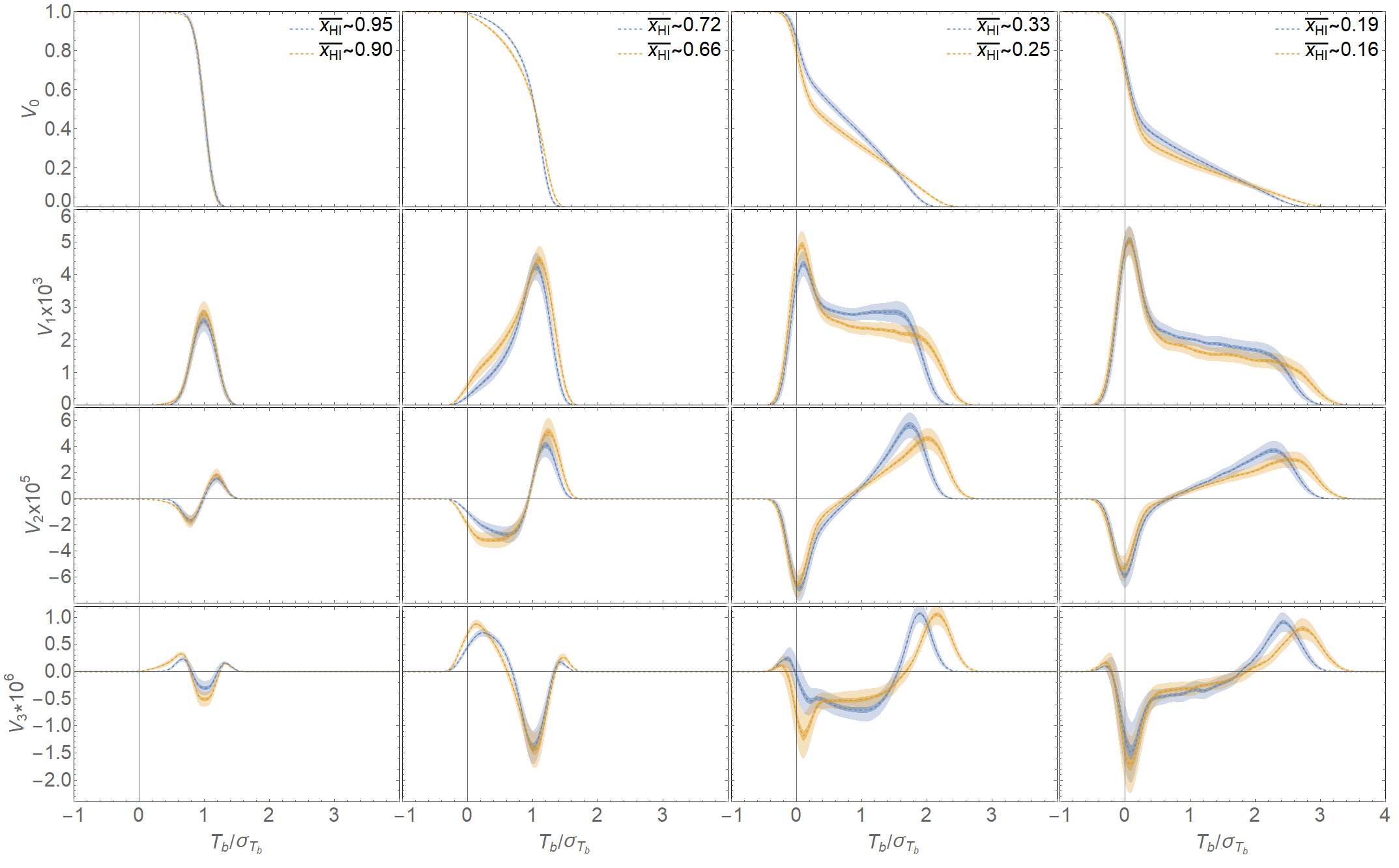}
\caption{The simulated observation of MFs and errors taken into account of the light cone effect 
at a few redshifts during EoR.  In each column we show the results of two mean neutral fractions
which are before (blue) and after (orange) a transition discussed in this paper. For each the lighter shadow shows the
error taken with a measurement of $(1\Gpc)^2 \times 48 \Mpc$, while
the darker shadow shows the error with a measurement of $(5\Gpc)^2 \times 48 \Mpc$. }
\label{MFs light cone effect} 
\end{figure*}

\subsection{The light cone effects}
In the above we have simulated the measurement of MFs at a given time or redshift in a comoving three dimensional box. 
However, in real observations one can only make the measurement along the light cone, the near and far sides
of the survey volume correspond to different redshifts. The size of the volume in which the measurement is made must then be
limited, such that the evolution is not too much during the light travel time across the box. 
For reionization, this poses a serious problem, because the ionization fraction are changing rapidly. 
If one take a redshift interval of $\delta z\sim0.2$ at $z=8.5$, for example, it corresponds to only 
about $\delta L\sim 50 \,\rm{Mpc}$ along the line of sight. 
As our smoothing length is $R_{\rm{pix}}=16 \,\rm{Mpc}$, a ``realistic'' observation can cover only about 
three slices. If such small volume is used, the uncertainties would be significant.  This light cone effect has 
been well known in the power spectrum measurement (e.g. \citealt{Datta:2011hv, Datta:2014fna}). 

Taking boxes of a thickness of $\delta L=48\Mpc$,  we calculate the MFs and the expected errors, the 
results are shown in Fig.~\ref{MFs light cone effect}. Here on each column 
we plot the results just before(in blue color) or after (in orange color) one of the transition of stages discussed above. 
The large lighter shadow depicts the expected error for a survey volume of $(1\Gpc)^2 \times 48 \Mpc$.
Due to the light cone effect, the error is much larger than the 
previously discussed sampling errors. Nevertheless, in few cases the differences are still large enough to be 
distinguished easily, but in other cases the differences are just comparable to the measurement errors. However, 
the measurement error can be reduced by observing a larger sky area. As noted earlier, 
the error of MFs roughly scales as $V^{-1/2}$. For a fixed redshift interval limited by the light cone effect, the 
error can be reduced to acceptable level by having a larger sky area. We also show in the darker shadow area in 
Fig.~\ref{MFs light cone effect} the result with a survey of $(5\Gpc)^2 \times 48 \Mpc$, with the same integration time per 
pixel. This would require a quite large amount of observing time, but may still be achieved over several years, and 
the process can be speed up if the SKA-low is equipped with multi-beam forming capability.

\section{Discussions}

We have analyzed the evolution of the topological structure of the 21 cm brightness temperature field during EoR 
using a set of semi-numerical simulations. The ionized regions have a brightness temperature of $T_b=0$, while the 
neutral regions have $T_b>0$. By tracking the  changes of MFs, especially by looking at the
MFs at just above $T_b=0$, one can gain insights on the topology of the ionization-neutral region interface during the
reionization process.  

Based on the analysis of the MFs and other statistics, e.g. the size distribution of the neutral and ionized regions,  
we find that the topological evolution of the reionization process can be divided into five different stages. 
In the first stage where the neutral fraction $x_{\rm HI} \gtrsim 0.9$, there are numerous isolated ionized regions (``bubbles"). 
As more numerous and larger bubbles formed, the ionized regions begin to connect with each other. 
Next, the ionized regions formed cluster structures (``ionized fibers"), with the largest one percolating 
through the whole simulation box.  At $x_{\rm HI} \approx 0.7$, most of the ionized regions overlap
with each other, forming a connected structure intertwined with a similar structure of neutral regions, the whole
space of ionized and neutral regions takes the appearance of a ``sponge". As the reionization progress further, the 
neutral regions begin to shrink and being surrounded by the ionized regions, at $x_{\rm HI} \lesssim 0.3$, a ``neutral fiber"
stage set in. Finally, the neutral fiber also shrinks and at some point around $x_{\rm HI} \sim 0.16$ it no longer
percolate through the whole simulation box, and we enter the stage of isolated ``neutral islands". 

From the results summarized above, we see that strictly speaking, the original assumptions of the bubble model and 
island model, namely that  the ionized regions or the neutral regions are isolated, are only valid during the 
first or the last stage of the reionization process. During the intermediate stages, the topology of the 
region could be fairly complicated, and as a result the numerical model may not be quantitatively accurate, as have been
noted in previous works \citep{Xu:2013npa,Furlanetto:2006jb}.
Nevertheless, it is interesting to note that the semi-numerical simulations which are based on spherical averages 
do produce very complex and irregular shaped regions. The bubble model and island model are based on the 
balance between the number of photons needed for ionization and the number of photons produced locally (bubble model) and 
also the number of photons from the external background (for the island model). 
The fiber-like shape in such models originates from the shape of the iso-density contours of the density 
fluctuations. Very recently, \citet{Bag:2018zon} used the MFs and cluster statistics to characterize the 
topology of the ionized regions, they  also found the fiber-like shapes 
of the ionized regions, and using the shapefinder statistics derived from the MFs, 
they found that the cross-section of the fibers are nearly constant. The numerical algorithm we adopted is not 
suitable for accurate computing of the shapefinder statistics, but this result is consistent with the general picture 
of reionization process we delineated above. 

Compared with the original assumption, the most important difference for the fiber shaped ionized region 
is that they will get some additional ionizing photons from the 
connected neighboring cells, i.e. some ionizing photons can propagate along the fibers, so that the ionization may be even 
faster. However, if the ionized regions are indeed shaped as fibers, the cross section for the background photon is fairly 
limited, and the zigzag path also limit the distance the photons could travel due to finite propagation time and the absorption by
the residue neutral gas in the ionized regions. Thus, even though the original assumption of isolated bubbles is no longer 
true during the ionized fiber stage,  we speculate that the model may still give reasonably good predictions.
For the neutral fibers, the inaccuracy of the 21cmFAST model may be larger, as a substantial amount of the photons could
come from the surrounding ionizing background. However, the IslandFAST program may take into account of the surface area
of the actual neutral region, the main difference in topology is that the neutral fibers are more prone to break-up into smaller
islands. 

We found that although the value of MFs are affected by thermal noise, which is more Gaussian than the signal and thus 
making the MFs more bland, there remains features which can be  used to distinguish the different stages, 
at least with high signal-to-noise measurements (${\rm SNR} \sim 10)$).
However, due to the rapid evolution of the ionization fraction during EoR, the light cone effect put a limit on the usable 
redshift bin size ($\Delta z \lesssim 0.2$). For a typical SKA EoR survey with an area of $\sim 40 \deg^2$ and 1000 hour integration 
time per pixel, in some cases the MFs can still be distinguished though for many cases the difference 
would be comparable to the measurement errors.  To achieve a  good measurement, a larger sky area is needed.   
Thus, the evolution of topology is within the reach through 21 cm brightness temperature measurements using SKA-low, 
though it may require a survey of many years to complete. The survey time could be shortened if the SKA-low
is equipped with multi-beam forming capability.

The present study is still limited in several ways. Our study is based on semi-numerical models, 
which is only an approximation and can not replace a full radiative transfer simulation. 
The study of the topology of reionization using the semi-numerical models 
may be considered as a self-consistent check on the basic assumptions of the method. 
We have limited our presentation to the results obtained for a fiducial model. The
different cosmological and reionization parameters may affect the results. However, from a few exploratory tests, 
it seems that as long as the ionizing sources are soft ones which produce clear boundary between the neutral and ionized  
regions, and the cosmological parameters and reionization parameters are not too extreme,
the general picture of the evolution presented above should be robust.

\acknowledgements
This work is supported by the Chinese Academy of Sciences (CAS) Frontier Science Key Project 
QYZDJ-SSW-SLH017, the National Natural Science Foundation of China (NSFC) key
project grant 11633004, the NSFC-ISF joint research program No. 11761141012, the CAS 
Strategic Priority Research Program XDA15020200, the National Key R\&D Program 
2017YFA0402603, 
the NSFC grant 11773034,  
and the MoST grant 2016YFE0100300. 
ZTC thanks Jiabei Zhu, Xianyun Jiang for valuable discussions.
 
\bibliography{references}
\bibliographystyle{hapj}
\end{document}